\documentclass[12pt,draftcls,onecolumn]{IEEEtran}

\usepackage{soul}
\usepackage{amsfonts}
\usepackage{dsfont}
\usepackage[cmex10]{amsmath}
\usepackage{amssymb,amsbsy,mathrsfs,amsthm}
\usepackage{amsmath,bm}

\usepackage{fancyhdr}

\usepackage{comment}

\usepackage{booktabs}  
\usepackage{array}         

\usepackage{mathptmx}
\usepackage{slashbox}
\usepackage{multirow}
\usepackage[table]{xcolor}
\usepackage{colortbl}

\usepackage{algorithm}
\usepackage{algpseudocode}
\algnewcommand{\algorithmicgoto}{\textbf{go to}}%
\algnewcommand{\Goto}[1]{\algorithmicgoto~\ref{#1}}%

\newtheorem{lemma}{Lemma}
\newtheorem{remark}{Remark}

\usepackage{graphicx}
\graphicspath{ {./} }
\DeclareGraphicsExtensions{.pdf}

\begin{document}

\title{Optimizing the Diffusion System Based on Continuous-Time Consensus Algorithm}

\author{Saber Jafarizadeh,~\IEEEmembership{Member,~IEEE,} \thanks{e-mail: saber.jafarizadeh@sydney.edu.au }   }

\maketitle


\begin{abstract}

Traditionally, systems governed by linear Partial Differential Equations (PDEs) are spatially discretized to exploit their algebraic structure and reduce the computational effort for controlling them.
Due to beneficial insights of the PDEs, recently, the reverse of this approach is implemented where a spatially-discrete system is approximated by a spatially-continuous one, governed by linear PDEs forming diffusion equations.
In the case of distributed consensus algorithms, this approach is adapted to enhance its convergence rate to the equilibrium.
In previous studies within this context, constant diffusion parameter is considered for obtaining the diffusion equations.
This is equivalent to assigning a constant weight to all edges of the underlying graph in the consensus algorithm.
Here, by relaxing this restricting assumption, a spatially-variable diffusion parameter is considered and by optimizing the obtained system, it is shown that significant improvements are achievable in terms of the convergence rate of the obtained spatially-continuous system.
As a result of approximation, the system is divided into two sections, namely, the spatially-continuous path branches and the lattice core which connects these branches at one end.
The optimized weights and diffusion parameter for each of these sections are optimal individually but considering the whole system, they are suboptimal.
It is shown that the symmetric star topology is an exception and the obtained results for this topology are globally optimal.
Furthermore, through variational method, the results obtained for the symmetric star topology are validated and it is shown that the variable diffusion parameter improves the robustness of the system too.

\end{abstract}

\begin{IEEEkeywords}
Distributed Consensus Algorithm, Diffusion System, Convergence Rate, Robustness, Variational Method 
\end{IEEEkeywords}


\section{Introduction}

In the context of distributed systems and control, the distributed consensus algorithm serves as the underlying mechanism for many other distributed algorithms developed for applications such as distributed estimation and detection for decentralized sensor networks, gossip algorithms \cite{Boyd06Gossip,AysalGossip13Application2009}, gossip algorithms \cite{Boyd06Gossip,AysalGossip13Application2009}, fastest mixing Markov chain problem \cite{Boyd11ConsensusApplication2004}, distributed data fusion in sensor networks \cite{OlshevskyConsensusApplication2006,KarConsensusApplication2007}, multiagent distributed coordination and flocking \cite{Gossip20Application2003,BlondelConsensusApplication2005}.
In this class of problems, the system is composed of subsystems referred to as agents.
Agents have the capability to exchange information in bidirectional manner with their neighbouring agents.
The underlying communication network topology is defined according to the neighborhood relation between agents.
Distributed consensus algorithm aims to reach a global consensus in each agent based on their initial states using only local interaction between agents.
See \cite{Gossip20Application2003,Gossip21Application2004} for an overview of distribute consensus algorithm.
One of the important design features of the distribute consensus algorithm is its convergence rate to the consensus state, which depends on the weights assigned to edges of the underlying network in the local updating procedure, see \cite{SaberQConsensusContinuous,SaberQuantGossip,SaberConsensusFusionTwoStars} for more details.
%

In the literature on the finite difference for PDE discretization \cite{Sarlette5Ref1974,EssersCourse2004,MortonBook2005,SarletteRef12}, systems involving linear partial differential equations with constant coefficient are spatially discretized in an effort to exploit their algebraic structure and reduce the computational effort of designing a controller.
By doing so, the spatial variables and their derivatives are mapped into the agent indexes and the links between them, respectively.
Authors in \cite{SarlettePDE2009} have taken the reverse of the approach in \cite{Sarlette5Ref1974,EssersCourse2004,MortonBook2005,SarletteRef12} and they have proposed a spatial-continuous approximation of distributed systems, including the consensus algorithm.
The main incentive for the approximation in \cite{SarlettePDE2009} is to benefit from intuitive insights of the PDE viewpoint and to derive more effective consensus protocols in terms of the convergence rate of the algorithm.
Based on the analysis and results in \cite{SarlettePDE2009}, authors in \cite{ItalyDiffusion2015} have proposed a linear local interaction strategy for the consensus algorithm and, they have generalized the average consensus algorithm to the infinite-dimensional setting of networked heat processes.
As a result, they have shown the eventual convergence of agents' states towards the spatial average of the agents' initial conditions.

The previous approach in the literature \cite{ItalyDiffusion2015,SarlettePDE2009} is to consider a constant diffusion parameter.
This is equivalent to assigning constant weight to all edges in the consensus algorithm.
In this paper, by considering a spatially-variable diffusion parameter and optimizing the obtained problem, we have shown that the convergence rate of 
the obtained diffusion system (the system of diffusion equations)
is improved compared to the case with constant diffusion parameter.
In our approach here, we have reached the diffusion equation in the continuum limit of the spatially-discrete and continuous-time consensus algorithm.
%
By doing so, the spatially-discrete path branches are transformed to spatially-continuous branches and as a result the network is divided into two parts, the spatially-continuous path branches and the lattice core which connects these branches at one end.
Using the optimal weights obtained in \cite{SaberQConsensusContinuous} for spatially-discrete continuous-time consensus algorithm, 
we have derived optimal weights for the lattice core of the network and the spatially-variable diffusion parameter for the path branches.
We have shown that the obtained optimal weights and diffusion parameter result in faster convergence rate compared to the spatially-constant diffusion parameter approach in \cite{ItalyDiffusion2015}.
An important issue regarding the results presented in this paper is that the weights obtained for the lattice core is optimal only for the lattice core individually, but considering the whole topology of the network, the obtained weights are suboptimal.
The only exception is the symmetric star topology, where the obtained results are globally optimal.
This is due to the fact that in the case of symmetric star topology, the lattice core is reduced to the central vertex in this topology.
Furthermore, through variational method we have validated the results obtained for the symmetric star topology.
By investigating the robustness of the diffusion system, 
we have shown that for the symmetric star topology, the robustness of the algorithm with variable diffusion parameter improves compared to the one with constant diffusion parameter.

The rest of this paper is organized as follows.
In section \ref{sec:ContinuumLimit}, the spatially-discrete continuous-time distributed consensus algorithm is formulated in the continuum limit.
Analysis of the resultant diffusion equations with constant and variable diffusion parameters are presented in Sections \ref{sec:ConstantDiffusionParameter} and \ref{sec:VariableDiffusionParameter}, respectively.
The special case of symmetric star topology is studied in Section \ref{SymmetricStarTopology} and section \ref{Conclusion} concludes the paper.

%

%
\section{Continuum Limit of Continuous-Time Consensus Algorithm}
\label{sec:ContinuumLimit}
In this section, we provide the derivation of the 
continuous-time distributed consensus algorithm in the continuum limit. 

We consider a network 
consisting of a given Lattice core (an arbitrary connected graph $\mathcal{G} = ( \mathcal{V} , \mathcal{E} ))$ where a path graph with $q$ vertices is connected to each one of the vertices in the lattice core.
We refer to the path graphs as tails.
%
%
%
We denote agents or vertices on branches by $(\alpha, j)$, for $\alpha \in \mathcal{G}$, $j=0,1,\ldots,q$  with corresponding agents' states $\boldsymbol{X}_{\alpha, j}$, $\alpha \in \mathcal{G}$, $j=0,1,\ldots,q$.
The weights on the edges of the Lattice core of the network are denoted by $\boldsymbol{W}_{\alpha\beta} = W_{(\alpha,0)(\beta,0)}$ for $\{\alpha,\beta\}\in \mathcal{E}(\mathcal{G})$ where $\mathcal{G}$ refers to the graph representing only the lattice core (excluding tails) and $\mathcal{E}$ is the set of edges in graph $\mathcal{G}$.
The weights on the edges of the path tails are denoted by $ \boldsymbol{W}_{i} = W_{(\alpha,i-1),(\alpha, i)}$ for  $i=1,2,\ldots, q$ and $\alpha \in \mathcal{G}$.
%
%
%
The state update equations of the continuous-time consensus algorithm \cite{SaberQConsensusContinuous} can be written as below,
\begin{subequations}
    \label{eq:StateUpdateEquationConsensusSpatialDiscrete265}
    \begin{gather}
        \frac{d}{dt}\boldsymbol{X}_{\alpha,0}(t)=\boldsymbol{W}_{1}{\left( \boldsymbol{X}_{\alpha,1}(t)-\boldsymbol{X}_{\alpha,0}(t)\right)}+\sum_{\{\alpha, \beta\}\in \mathcal{E}(\mathcal{G})}\boldsymbol{W}_{\alpha\beta}{\left(\boldsymbol{X}_{\beta,0}(t)-\boldsymbol{X}_{\alpha,0}(t)\right)},  \label{eq:StateUpdateEquationConsensusSpatialDiscrete265a} \\ 
        \frac{d}{dt}\boldsymbol{X}_{\alpha,j}(t) = \boldsymbol{W}_{j}{\left( \boldsymbol{X}_{\alpha,j-1}(t) - \boldsymbol{X}_{\alpha,j}(t) \right) } + \boldsymbol{W}_{j+1} { \left( \boldsymbol{X}_{\alpha,j+1}(t) - \boldsymbol{X}_{\alpha,j}(t) \right) }, \label{eq:StateUpdateEquationConsensusSpatialDiscrete265b} \\  
        \frac{d}{dt}\boldsymbol{X}_{\alpha,q}(t) = \boldsymbol{W}_{q} { \left( \boldsymbol{X}_{\alpha,q-1}(t) - \boldsymbol{X}_{\alpha,q}(t) \right) },  \label{eq:StateUpdateEquationConsensusSpatialDiscrete265c}
    \end{gather}
\end{subequations}
where (\ref{eq:StateUpdateEquationConsensusSpatialDiscrete265b}) holds for $j=1,2,\ldots ,q-1$.
In the continuum limit of $q\rightarrow \infty$, with the constraint that $\lim_{ q \rightarrow \infty  } \frac{j}{q} = \xi$ is finite, we have $d\xi=\frac{1}{q}$.
We use $Q_{\alpha}(\xi,t)$ to denote the state of the agents, i.e. $Q_{\alpha}(\xi,t)=X_{\alpha,j}(t)$, $Q_{\alpha,0}(t)=X_{\alpha,0}(t)$.
In the continuum limit, 
the path tails are transformed into bar tails and the state update equation (\ref{eq:StateUpdateEquationConsensusSpatialDiscrete265b}) can be written as below,
\begin{equation}
    \label{eq:ContinuousConsensusStateEvolution281}
    \begin{aligned}
        &\frac {\partial} {\partial t} \boldsymbol{Q}_{\alpha} (\xi,t)  =
        \boldsymbol{W}(\xi + d\xi) \left( \boldsymbol{Q}_{\alpha} (\xi+d\xi,t) -  \boldsymbol{Q}_{\alpha} (\xi , t)  \right)
        -
        \boldsymbol{W}( \xi ) \left( \boldsymbol{Q}_{\alpha} (\xi,t) -  \boldsymbol{Q}_{\alpha} (\xi-d\xi , t)  \right)
        = \\
        &\left. \left(  \frac{\boldsymbol{W}(\nu)}{q} \cdot \frac{\partial}{\partial \nu} \boldsymbol{Q}_{\alpha}  (\nu,t)  \right) \right|_{ \nu = \xi + d\xi }
        -
        \left. \left(  \frac{\boldsymbol{W}(\nu)}{q} \cdot \frac{\partial}{\partial \nu} \boldsymbol{Q}_{\alpha}  (\nu,t)  \right) \right|_{ \nu = \xi }
        = 
        %
        %
        \frac{1}{q^2} \frac{\partial}{\partial \xi} \left(  \boldsymbol{W}(\xi)  \frac{\partial}{\partial \xi} \boldsymbol{Q}_{\alpha} \left( \xi , t \right)  \right).
    \end{aligned}
\end{equation}

\section{Constant Diffusion Parameter}
\label{sec:ConstantDiffusionParameter}

In this section, we analyse the diffusion equation (\ref{eq:ContinuousConsensusStateEvolution281}) obtained from modelling the continuous-time consensus algorithm in the continuum limit where the diffusion parameter $(\hat{\Theta})$ is constant all over the network.
The main focus of the analysis presented in this section is to formulate (and thus optimize) the convergence rate of the agents' states to their equilibrium point.
To properly model the continuous-time consensus algorithm in the continuum limit as diffusion equations with constant diffusion parameter $\hat{\Theta}$, 
the weights on the edges of the lattice $(\mathcal{W}_{\alpha\beta})$ and the diffusion coefficient $(\hat{\Theta})$ should be selected in accordance with the following equations,
%
%
%
%
\begin{subequations}
      \label{eq:WeightsContinuumConstantTheta311}
    \begin{gather}
        \boldsymbol{W}_{j} = q^2 \cdot \hat{\Theta} \quad \text{for} \quad j=1,2,\ldots, q \label{eq:WeightsContinuumConstantTheta311a}\\
        \boldsymbol{W}_{\alpha\beta}=q \cdot \hat{\Theta} \cdot \mathcal{W}_{\alpha\beta} \quad \text{for}\quad \{\alpha, \beta\}\in \mathcal{E}(\mathcal{G}) \label{eq:WeightsContinuumConstantTheta311b}
    \end{gather}
\end{subequations}
Note that in case of continuous-time consensus algorithm with uniform weights, the weights on edges of the path tails $(\boldsymbol{W}_{j})$ are equal to each other.
Using the weights in (\ref{eq:WeightsContinuumConstantTheta311}), the state update equations (\ref{eq:StateUpdateEquationConsensusSpatialDiscrete265}) in the continuum limit can be written as below,
%
%
\begin{subequations}
    \label{eq:StateUpdateEquationsContinuumConstantTheta323}
    \begin{gather}
        \frac{\partial}{\partial t}Q_{\alpha}(\xi,t) = \hat{\Theta} \cdot \frac{\partial^2}{\partial\xi^2}Q_{\alpha}(\xi,t), \quad \text{for} \quad \alpha \in \mathcal{G}   \label{eq:StateUpdateEquationsContinuumConstantTheta323a}  \\
        \frac{\partial}{\partial t} Q_{\alpha}(1,t) = -\hat{\Theta} \cdot q \cdot \frac{\partial}{\partial\xi} Q_{\alpha}(\xi,t)\large{|}_{\xi=1}, \quad \text{for} \quad \alpha \in \mathcal{G}         \label{eq:StateUpdateEquationsContinuumConstantTheta323b} \\
       \frac{\partial}{\partial t}Q_{\alpha}(0,t) = \hat{\Theta} \cdot q \cdot {\left(  \frac{\partial}{\partial\xi}Q_{\alpha}(\xi,t)|_{\xi=0}+\sum_{\beta \in \mathcal{N}(\alpha)}\mathcal{W}_{\alpha\beta}{\left(Q_{\beta}(0,t)-Q_{\alpha}(0,t)\right)}\right)}  \label{eq:StateUpdateEquationsContinuumConstantTheta323c}
    \end{gather}
\end{subequations}
%
To have finite values for $\frac{\partial}{\partial t}Q_{\alpha}(1,t)$ and $\frac{\partial}{\partial t}Q_{\alpha}(0,t)$, it is required that $\frac{\partial}{\partial\xi}Q_{\alpha}(\xi,t)|_{\xi=1} = 0$ and $\frac{\partial}{\partial\xi}Q_{\alpha}(\xi,t)|_{\xi=0}+\sum_{\beta \in \mathcal{N}(\alpha)} \mathcal{W}_{\alpha\beta} {\left(Q_{\beta}(0,t) - Q_{\alpha}(0,t)\right)} = 0$, respectively.
Thus the diffusion equation (\ref{eq:StateUpdateEquationsContinuumConstantTheta323a}) and the boundary conditions (\ref{eq:StateUpdateEquationsContinuumConstantTheta323b}) and (\ref{eq:StateUpdateEquationsContinuumConstantTheta323c}) can be written as below,
%
%
%
%
%
\begin{subequations}
    \label{eq:StateUpdateEquationsContinuumConstantThetaFinalForm344}
    \begin{gather}
        \frac{\partial}{\partial t}Q_{\alpha}(\xi,t) = \hat{\Theta} \frac{\partial^2}{\partial\xi^2}Q_{\alpha}(\xi,t), \quad \text{for} \quad \alpha \in \mathcal{G}   \label{eq:StateUpdateEquationsContinuumConstantThetaFinalForm344a}  \\
        \frac{\partial}{\partial\xi}Q_{\alpha}(\xi,t)\big{|}_{\xi=1}=0 \quad \text{for} \quad \alpha \in \mathcal{G}  \label{eq:StateUpdateEquationsContinuumConstantThetaFinalForm344b}  \\
        \frac{\partial}{\partial\xi}Q_{\alpha}(\xi,t)\big{|}_{\xi=0}+\sum_{\beta \in \mathcal{N}(\alpha)}\mathcal{W}_{\alpha\beta} \cdot {\left(Q_{\beta}(0,t) - Q_{\alpha}(0,t)\right)} = 0  \quad \text{for} \quad \alpha \in \mathcal{G}  \label{eq:StateUpdateEquationsContinuumConstantThetaFinalForm344c}
    \end{gather}
\end{subequations}
Note that the boundary condition (\ref{eq:StateUpdateEquationsContinuumConstantThetaFinalForm344b}) is of Neumann-type while the boundary condition (\ref{eq:StateUpdateEquationsContinuumConstantThetaFinalForm344c}) is of Robin-type.


There are major differences between the diffusion equation and the boundary condition in (\ref{eq:StateUpdateEquationsContinuumConstantThetaFinalForm344}) and those presented in \cite{ItalyDiffusion2015}.
In \cite{ItalyDiffusion2015}, the weights on the edges of the Lattice graph $\mathcal{G}$ are assumed to be constant and equal to one while here in this paper, this limitation is relaxed and it is assumed that these weights can have different values.
Another difference is the alignment of the path bars.
In \cite{ItalyDiffusion2015}, the path bars are coupled together at $\xi = 1$ and their other end $(\xi = 0)$ is free.
While in this paper the alignment of the path bars is reverse of \cite{ItalyDiffusion2015}, i.e. the path bars are connected to the lattice at $\xi = 0$ and their other end $(\xi = 1)$ is free.
As a result, the diffusion equation obtained in \cite{ItalyDiffusion2015} has a negative sign while in the diffusion equation (\ref{eq:StateUpdateEquationsContinuumConstantThetaFinalForm344a}) there is no negative sign.



%
We define the vector $\boldsymbol{Q}(\xi,t) = \left[  Q_{1}(\xi,t), \ldots, Q_{N}(\xi,t)  \right]^{T}$ as the vector representing the state of 
the diffusion system
where $N = |\mathcal{G}|$. 
From (\ref{eq:StateUpdateEquationsContinuumConstantThetaFinalForm344}) it is obvious that the 
dynamics of the diffusion system
evolve according to the following diffusion equation
\begin{equation}
    \label{eq:DiffusionPDEStateUpdate229}
    \begin{aligned}
        \frac{\partial}{\partial t} \boldsymbol{Q}(\xi,t)  =  \hat{\Theta} \cdot \frac{\partial^{2}}{\partial \xi^{2}}\boldsymbol{Q}(\xi,t),
    \end{aligned}
\end{equation}
%
with the following boundary conditions,
\begin{subequations}
    \label{eq:BoundaryConditions404}
    \begin{gather}
        \frac{\partial}{\partial \xi}\boldsymbol{Q}(\xi,t)|_{\xi = 1} = 0, 		\label{eq:BoundaryConditions240a}\\
         \frac{\partial}{\partial \xi}\boldsymbol{Q}(\xi,t)|_{\xi = 0} = \boldsymbol{L}_{w} \times \boldsymbol{Q}(0,t), \label{eq:BoundaryConditions240b} 
    \end{gather}
\end{subequations}
where $\boldsymbol{L_{w}}$ is the weighted Laplacian matrix of the lattice graph $\mathcal{G}$.
%
%
The boundary condition (\ref{eq:BoundaryConditions240b}) is based on the local interaction protocol proposed in \cite{ItalyDiffusion2015}.
Authors in \cite{ItalyDiffusion2015} have considered the unweighted Laplacian matrix $(\boldsymbol{L})$ of the lattice graph $\mathcal{G}$ and they have shown that the closed-loop system is stable in the space $H^2(0,1)$ and system eventually reaches the average consensus given as below,
\begin{equation}
    \label{eq:ConsensusPoint260}
    \begin{aligned}
        \lim_{t \rightarrow \infty} {  \boldsymbol{Q}(\xi,t)  }  =  \left(   \frac{1}{N} \int_{0}^{1} { \boldsymbol{1}^{T} \boldsymbol{Q}(\xi,0) d\xi  } \right) \cdot \boldsymbol{1}, \quad \forall \xi \in (0,1).
    \end{aligned}
\end{equation}
$\boldsymbol{Q}(\xi,0)$ is the given initial state of the system and $\boldsymbol{1}$ is a column vector of size $N$ with all elements equal to one.
Note that the term $\left(   \frac{1}{N} \int_{0}^{1} { \boldsymbol{1}^{T} \boldsymbol{Q}(\xi,t) d\xi  } \right)$ is the spatial averaging of the agents initial conditions $\left( \boldsymbol{Q}(\xi,t) \right)$.


The main objective of the study presented here is to optimize the convergence rate of the vector of states $\boldsymbol{Q}(\xi,t)$ to its consensus equilibrium point (\ref{eq:ConsensusPoint260}).
%
%
Here in this paper, we use weighted Laplacian matrix $(\boldsymbol{L}_{w})$ instead of the unweighted Laplacian matrix $(\boldsymbol{L})$ in order to improve the convergence rate of the state vector $\boldsymbol{Q}(\xi,t)$ to its equilibrium state (\ref{eq:ConsensusPoint260}).
$\boldsymbol{L}_{w}$ is a symmetric matrix and based on the SVD decomposition it can be written as below,
\begin{equation}
    \label{eq:LaplacianDecomposition288}
    \begin{aligned}
        \boldsymbol{L}_{w}  =  \sum_{k=2}^{N} { \lambda_k \boldsymbol{\eta}_{k} \boldsymbol{\eta}_{k}^{T} },
    \end{aligned}
\end{equation}
where $\lambda_k$ and $\boldsymbol{\eta}_{k}$ are the eigenvalues and eigenvectors of $\boldsymbol{L}_{w}$, respectively.
Note that the first eigenvalue of $\boldsymbol{L}_{w}$ is zero i.e. $\boldsymbol{L}_{w} \times \boldsymbol{\eta}_{1} = 0$.
The eigenvalues of the weighted Laplacian matrix $\left( \boldsymbol{L}_{w} \right)$ can be sorted as below,
\begin{equation}
    \label{eq:LaplacianEigenvalues327}
    \begin{aligned}
        \lambda_{N}\left( \boldsymbol{L}_{w} \right)
        \geq
        \cdots
        \geq
        \lambda_{2}\left( \boldsymbol{L}_{w} \right)
        >
        \lambda_{1}\left( \boldsymbol{L}_{w} \right)
        = 0.
    \end{aligned}
\end{equation}
\begin{lemma}
\label{lemma1}
For a given weighted Laplacian matrix $\left( \boldsymbol{L}_{w} \right)$ with eigenvalues sorted as in (\ref{eq:LaplacianEigenvalues327}), the state vector $\boldsymbol{Q}(\xi,t)$ of the diffusion system (\ref{eq:DiffusionPDEStateUpdate229}) with constant diffusion parameter $\hat{\Theta}$ can be written as below,
\begin{equation}
    \label{eq:539Uniform}
    \begin{aligned}
        &
        \boldsymbol{Q}(\xi,t)
        =
        \left(
        \left( \frac{1}{\sqrt{N}} \cdot \int_{0}^{1} { \boldsymbol{1}^{T} \times \boldsymbol{Q} (\xi^{'},0) d\xi^{'}  } \right)   + 
        \sum_{ n_{1} = 1 }^{\infty} {    A_{1,n_{1}}     \cdot e^{- \mu_{1,n_{1}} \cdot t} \cdot \cos{ \left( n_{1} \cdot \pi \cdot \xi \right) }  }
        \right)
        \cdot \frac{1}{\sqrt{N}} \boldsymbol{1}
        + \\
        %
        %
        &
        \qquad\qquad
        \sum_{k=2}^{N} {
        \left(
        \sum_{ n_{k} = 1 }^{\infty}  A_{k,n_{k}}   \cdot   e^{ -\mu_{k,n_{k}} \cdot t }  \cdot \cos{\left(  \sqrt{\frac{\mu_{k,n_{k}}}{\hat{\Theta}}} \cdot (1-\xi)  \right)}
        \right)
        \cdot \boldsymbol{\eta}_{k}
        }
    \end{aligned}
\end{equation}
where $A_{1,n_{1}} = \frac{2}{\sqrt{N}} \cdot \left( \int_{0}^{1} { \left( \boldsymbol{1}^{T} \times \boldsymbol{Q}(\xi,0) \right) \cdot \cos{\left(n_1 \cdot \pi \cdot \xi\right)} \cdot d\xi  } \right)$
and
$A_{k,n_{k}} = \left(     2 \lambda_{k} / \left( \lambda_{k} + \sin^{2}{\left( \mu_{k,n_{k}}/\hat{\Theta} \right)} \right)    \right)$  $\cdot$  $\left( \int_{0}^{1} \right.$   $ \left( \boldsymbol{\eta}_{k}^{T}  \times \boldsymbol{Q}(\xi,0) \right) \cdot \cos{ \left(  \sqrt{  \mu_{k,n_{k}} / \hat{\Theta}  } \cdot (1-\xi)  \right) } \cdot d\xi     \Big)$,
and $\mu_{1,n_{1}} = n_{1}^{2} \cdot \pi^2 \cdot \hat{\Theta}$ for $n_{1} = 1,\ldots,\infty$ and $\mu_{k,n_{k}}$ for $k=2,\ldots,N$ and $n_{k} = 1,\ldots,\infty$ are obtained from the roots of the following equation,
\begin{equation}
    \label{eq:KthMuAnswerUniform558}
    \begin{aligned}
        \sqrt{  \frac{\mu_{k,n_{k}}}{\hat{\Theta}}  }  =  \lambda_{k}  \cdot \cot{ \left(  \sqrt{ \frac{\mu_{k,n_{k}}}{\hat{\Theta}} }  \right) },
    \end{aligned}
\end{equation}
\end{lemma}
Proof of this lemma is provided in Appendix \ref{sec:Proof1}.
\begin{remark}
\label{remark1}
From (\ref{eq:539Uniform}) it can be concluded that the convergence rate of the state vector $\boldsymbol{Q}(\xi,t)$ to its consensus equilibrium point (\ref{eq:ConsensusPoint260}) is governed by the sentence that includes the smallest of  $\mu_{k,n_{k}}$.
All $\mu_{k,n_{k}}$ have real positive values since the underlying graph of the network is an undirected graph.
$\mu_{2,1}$ is the smallest of  $\mu_{k,n_{k}}$.
Therefore, the convergence rate of the state vector $\boldsymbol{Q}(\xi,t)$ to its consensus equilibrium point (\ref{eq:ConsensusPoint260}) is governed by $\mu_{2,1}$.
%
If we consider the diffusion equation as a dynamical system then $\mu_{k,n_{k}}$ are the Lyapunov exponents of the system and their inverse i.e. $1 / \mu_{k,n_{k}}$ are the relaxation times of the system.
%
%
%
%
%
\end{remark}

\subsection*{Convergence Rate Using Constant Diffusion Parameter}
\label{sec:ConvergenceRateUniform}

\begin{figure}
  \centering
     \includegraphics[width=100mm]{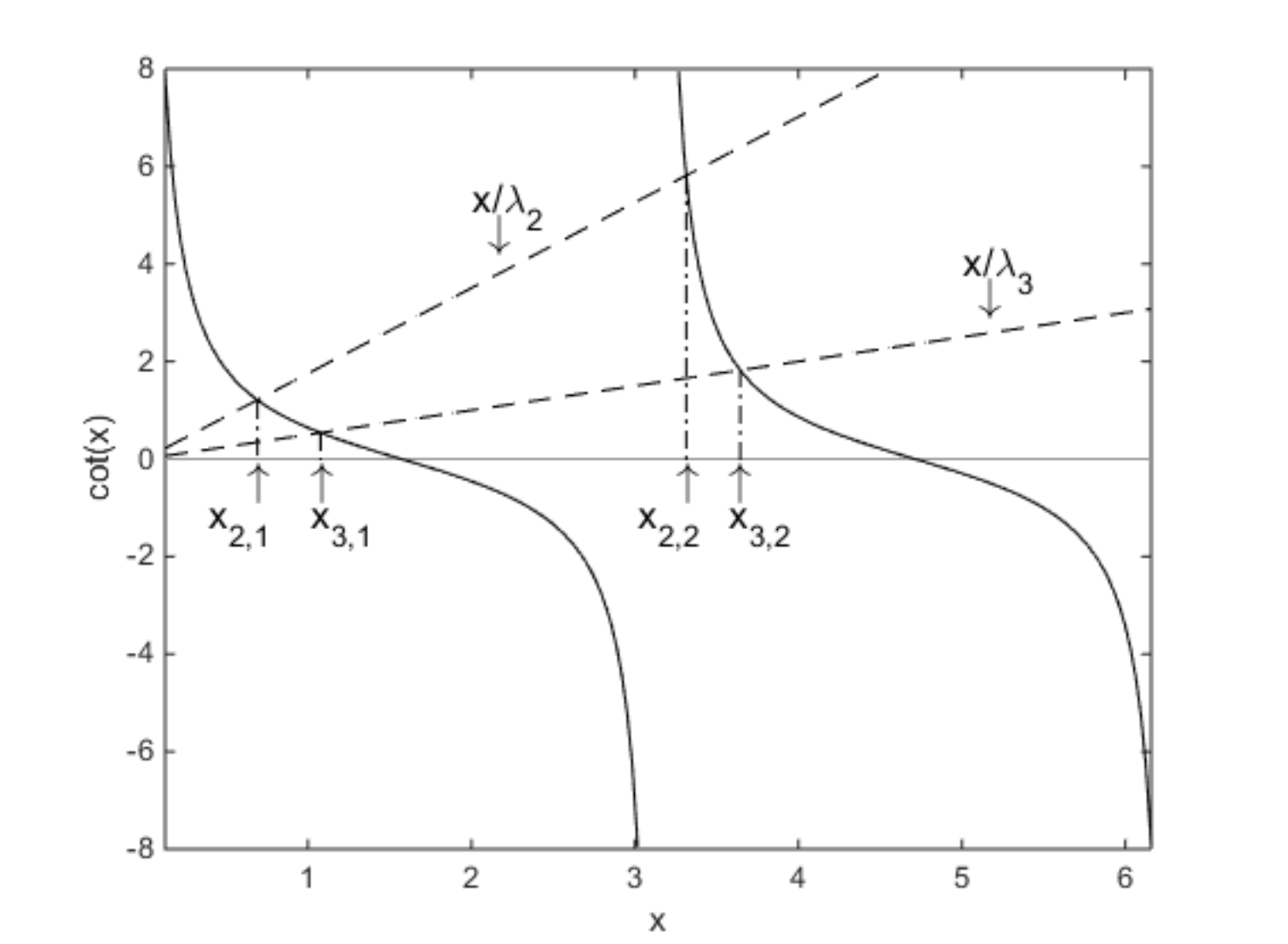}
  \caption{Roots of equation (\ref{eq:MuEquation550}).}
  \label{fig:Figure1CotRoots}
\end{figure}

In this subsection, we aim to address the convergence rate of the state vector $\boldsymbol{Q}(\xi,t)$ to its consensus equilibrium point (\ref{eq:ConsensusPoint260}).
%
%
As stated in remark \ref{remark1}, the convergence rate of the state vector $\boldsymbol{Q}(\xi,t)$ to its consensus equilibrium point (\ref{eq:ConsensusPoint260}) is governed by the smallest positive Lyapunov exponent, i.e. $\mu_{2,1}$.
To further explain the statement in remark \ref{remark1},
by defining variable $ x_{k,n_{k}} = \sqrt{ \frac{\mu_{k,n_{k}}}{\hat{\Theta}} } $, equation (\ref{eq:KthMuAnswerUniform558}) can be written as below,
\begin{equation}
    \label{eq:MuEquation550}
    \begin{aligned}
        x_{k,n_{k}} = \lambda_{k}  \cdot  \cot{ \left( x_{k,n_{k}} \right) }.
    \end{aligned}
\end{equation}
The roots of equation (\ref{eq:MuEquation550}) is visualized in Figure \ref{fig:Figure1CotRoots}.
As an example in Figure \ref{fig:Figure1CotRoots}, it is shown that intersection of the lines $x / \lambda_{2}$ and $x / \lambda_{3}$ with function $\cot{(x)}$ results in the roots $x_{2,1}$, $x_{2,2}$ and $x_{3,1}$, $x_{3,2}$ respectively.
Note that since $\lambda_{1} =0$, the line $x / \lambda_{1}$ is a vertical line that intersects with the function $\cot{(x)}$ at points $n  \cdot \pi$ where both functions reach infinity.
It is obvious from Figure \ref{fig:Figure1CotRoots} that $x_{2,1}$ 
is the smallest positive root of (\ref{eq:MuEquation550}).
Therefore, for $\mu_{2,1}$ we have
\begin{equation}
    \label{eq:Mu2Equation661}
    \begin{aligned}
        \mu_{2,1} = \hat{\Theta}  \cdot  x_{2,1}^{2},
    \end{aligned}
\end{equation}
and we can conclude that the convergence rate of the state vector $\boldsymbol{Q}(\xi,t)$ to its consensus equilibrium point (\ref{eq:ConsensusPoint260}) is governed by $\mu_{2,1}$.
Thus to optimize the convergence rate, the second smallest eigenvalue of the weighted laplacian matrix (i.e. $\lambda_{2}(\boldsymbol{L}_{w})$) has to be maximized.
For a given lattice graph $\mathcal{G} = \left( \mathcal{V} , \mathcal{E} \right)$, this optimization problem can be written as below,
\begin{equation}
    \label{eq:OptimizationProblemInitial590}
    \begin{aligned}
        \max\limits_{\boldsymbol{w}} \quad &\lambda_{2}(\boldsymbol{L}_{w}) \\
        s.t. \quad &\sum\limits_{ \{i,j\} \in \mathcal{E} } { \boldsymbol{w}_{ij} }  \leq  D_L,
    \end{aligned}
\end{equation}
where $\boldsymbol{w}_{ij}$ for $\{i,j\} \in \mathcal{E}$ are the weights assigned to the weighted Laplacian matrix $\boldsymbol{L}_{w}$.
$D_{L}$ is the upper limit on sum of the weights on edges of the lattice core.

This optimization problem is similar to the Classical Continuous Time Consensus problem as described in \cite{SaberQConsensusContinuous}.
In the following, we have provided the optimal weights and the resultant value of the convergence rate $(\mu_{2,1})$ obtained from (\ref{eq:MuEquation550}) for different topologies.
The results presented here are based on those provided in \cite{SaberQConsensusContinuous}.
Note that, in general providing closed-form formula for the root $x_{2,1}$ and therefore the convergence rate is not feasible, since equation (\ref{eq:MuEquation550}) should be solved numerically for each specific topology.

\subsubsection{Topologies with $N=2, 3$ \& $4$ Vertices}
\begin{figure}
  \centering
     \includegraphics[width=100mm]{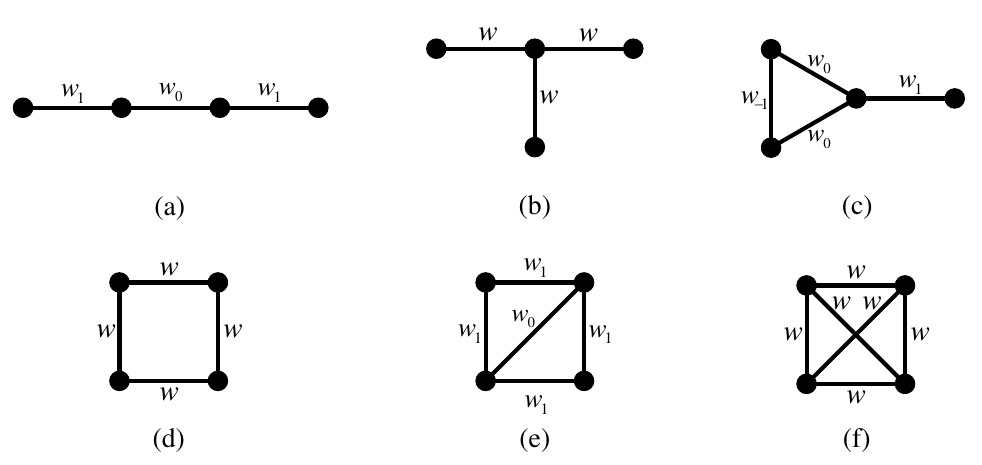}
  \caption{All possible connected topologies with N = 4 vertices.}
  \label{fig:N4Graphs}
\end{figure}

For a network with $N = 2$ vertices, the only connected topology is the path graph with $2$ vertices.
The optimal value of the second smallest eigenvalue $(\lambda_{2}(\boldsymbol{L}_{w}))$ for path topology with $2$ vertices is $2$ and the optimal weight is $1$.
The optimal value of the root $x_{2,1}$ for this topology is $1.0768$.
For $N = 3$ vertices, there are two connected topologies, namely, path topology and the triangular topology which is a complete graph.
In case of the path topology with $3$ vertices, the optimal value of the second smallest eigenvalue of the weighted Laplacian matrix $(\lambda_{2}(\boldsymbol{L}_{w}))$ is $1$ and the optimal weight is $1$.
The optimal value of the 
root $x_{2,1}$ for this topology is $0.8603$.
In case of the triangular topology, the optimal values of $\lambda_{2}(\boldsymbol{L}_{w})$, weight and the 
root $x_{2,1}$ are $3$, $1$ and $1.1924$, respectively.
There are six connected topologies with $N=4$ vertices.
These topologies are depicted in Figure \ref{fig:N4Graphs}.
The optimal results for these topologies are provided in Table \ref{tab:DiffusionTableN4}.
\begin{table}
\centering
    \caption{The optimal weights and the convergence rate $(\mu_{2,1})$ for all possible connected topologies with $N = 4$ vertices.The results presented in this table are obtained for $\hat{\Theta}=1$.}
    %
    \label{tab:DiffusionTableN4}
    \begin{tabular}{|c|c|c|c|c|c|} \hline
    {\multirow{2}{*}{Topology}}                  & {\multirow{2}{*}{Weights}}                             & \multicolumn{2}{|c|}{Constant $\Theta$}      & \multicolumn{2}{|c|}{Variable $\Theta$}       \\ \cline{3-6}
    &{}          &{$\mu_{2,1}$ for $D_L=|\mathcal{V}|$}     &{$\mu_{2,1}$ for $D_L=|\mathcal{E}|$}      &{$\mu_{2,1}$ for $D_L=|\mathcal{V}|$}     &{$\mu_{2,1}$ for $D_L=|\mathcal{E}|$}     \\ \hline
    {\multirow{2}{*}{Path}}     & {$w_{0}=2D_L/5$}                         & {\multirow{2}{*}{$0.6257$}}                &{\multirow{2}{*}{$0.4971$}}    &{\multirow{2}{*}{$0.9026$}}     &{\multirow{2}{*}{$0.7236$}} \\
    {}                          & {$w_{1}=3D_L/10$}                        & {}    &{} & {}    &{} \\ \hline
    {Star}                      & {$w = D_L/3$}                             & {$0.9047$}        &{$0.7402$}  & {$1.2772$}        &{$1.0586$}  \\ \hline
    {\multirow{3}{*}{Lollipop}} & {$w_{-1}  =  D_L( 2 - \sqrt{3} )/6$}      & {\multirow{3}{*}{$1.0517$}} & {\multirow{3}{*}{$1.0517$}} & {\multirow{3}{*}{$1.4668$}} & {\multirow{3}{*}{$1.4668$}}  \\
    {}                          & {$w_{0} = D_L/3$}                       & {}    & {}   & {}    & {}  \\
    {}                          & {$w_{1} = D_L/2$}                         & {}    & {}   & {}    & {}  \\ \hline
    {Cycle}                     & {$w = D_L/4$}                             & {$1.1597$}    & {$1.1597$}  & {$1.6022$}    & {$1.6022$}  \\ \hline
    {\multirow{2}{*}{Paw}}      & {$w_0 = 0$}                           & {\multirow{2}{*}{$1.1597$}}  & {\multirow{2}{*}{$1.3047$}}   & {\multirow{2}{*}{$1.6022$}}  & {\multirow{2}{*}{$1.7792$}} \\
    {}                          & {$w_1 = D_L/4$}                         & {}        & {}  & {}        & {}  \\ \hline
    {Complete Graph}            & {$w = D_L/6$}                             & {$1.3465$}        & {$1.5992$} & {$1.8295$}        & {$2.1215$} \\  \hline
    \end{tabular}
\end{table}

\subsubsection{Complete Graph}
Complete graph is a topology that all vertices are connected to each other.
In a complete graph with $N$ vertices, there are $N(N-1)/2$ edges.
For this topology the optimal weight on all edges is $w = 2D_L / \left(N (N-1) \right)$ and the optimal value of $\lambda_{2}(\boldsymbol{L}_{w})$ is equal to $2D_L/(N-1)$, where $D_L$ is the upper limit on sum of the weights on edges of the lattice core.
The optimal value of $\mu_{2,1}$ for different values of $N$ is provided in Tables 
\ref{tab:NumericalResultsDVertices} and \ref{tab:NumericalResultsDEdges}.

\begin{table}
\centering
    \caption{The optimal Convergence Rate $(\mu_{2,1})$ using constant and variable diffusion parameter for complete graph, path and cycle topologies with $N$ vertices. Ratio is the value of $\mu_{2,1}$ obtained for variable diffusion parameter divided by that of the constant parameter. The results presented in this table are obtained for $\hat{\Theta} = 1$ and $D_L=N$ (i.e. the upper limit on sum of the weights on the edges of the lattice core $(D_L)$ is set equal to the number of vertices $(N)$ in the graph.)}
    %
    \label{tab:NumericalResultsDVertices}
    \begin{tabular}{|c|c|c|c|c|c|c|c|c|c|} \hline
    \multirow{2}{*}{N}                   & \multicolumn{3}{|c|}{Complete Graph}        & \multicolumn{3}{|c|}{Path}            & \multicolumn{3}{|c|}{Cycle}     \\ \cline{2-10}
    {}                    & {Constant $\Theta$}     & {Variable $\Theta$} &{Ratio}  & {Constant $\Theta$}     & {Variable $\Theta$} &{Ratio}  & {Constant $\Theta$}     & {Variable $\Theta$}  &{Ratio}   \\ \hline
    {5}   &{$1.3047$} &{$1.7792$} &{$1.3637$}   &{$0.4268$} &{$0.6243$} &{$1.4627$}   &{$0.9263$} &{$1.3053$} &{$1.4092$}  \\ \hline
    {6}   &{$1.2782$} &{$1.7474$} &{$1.3671$}   &{$0.3070$} &{$0.4526$} &{$1.4743$}   &{$0.7402$} &{$1.0585$} &{$1.4300$}  \\ \hline
    {7}   &{$1.2598$} &{$1.7253$} &{$1.3695$}   &{$0.2305$} &{$0.3414$} &{$1.4811$}   &{$0.5969$} &{$0.8627$} &{$1.4453$}  \\ \hline
    {8}   &{$1.2464$} &{$1.7090$} &{$1.3711$}   &{$0.1790$} &{$0.2660$} &{$1.4860$}   &{$0.4874$} &{$0.7100$} &{$1.4567$}  \\ \hline
    {9}   &{$1.2362$} &{$1.6963$} &{$1.3722$}   &{$0.1428$} &{$0.2126$} &{$1.4888$}   &{$0.4033$} &{$0.5907$} &{$1.4647$}  \\ \hline
    {10}  &{$1.2281$} &{$1.6864$} &{$1.3732$}   &{$0.1165$} &{$0.1737$} &{$1.4910$}   &{$0.3379$} &{$0.4971$} &{$1.4711$}  \\ \hline
    {11}  &{$1.2216$} &{$1.6787$} &{$1.3742$}   &{$0.0968$} &{$0.1445$} &{$1.4928$}   &{$0.2866$} &{$0.4230$} &{$1.4759$}  \\ \hline
    {12}  &{$1.2162$} &{$1.6720$} &{$1.3748$}   &{$0.0816$} &{$0.1220$} &{$1.4951$}   &{$0.2456$} &{$0.3633$} &{$1.4792$}  \\ \hline
    {13}  &{$1.2116$} &{$1.6664$} &{$1.3754$}   &{$0.0698$} &{$0.1043$} &{$1.4943$}   &{$0.2126$} &{$0.3151$} &{$1.4821$}  \\ \hline
    {14}  &{$1.2078$} &{$1.6615$} &{$1.3756$}   &{$0.0603$} &{$0.0902$} &{$1.4959$}   &{$0.1857$} &{$0.2756$} &{$1.4841$}  \\  \hline
    \end{tabular}
\end{table}

\begin{table}
\centering
    \caption{The optimal Convergence Rate $(\mu_{2,1})$ using constant and variable diffusion parameter for complete graph, path and cycle topologies with $N$ vertices. Ratio is the value of $\mu_{2,1}$ obtained for variable diffusion parameter divided by that of the constant parameter. The results presented in this table are obtained for $\hat{\Theta} = 1$ and the number of edges in the graph as the upper limit on sum of the weights on edge of the lattice core $(D_L)$.}
    %
    \label{tab:NumericalResultsDEdges}
    \begin{tabular}{|c|c|c|c|c|c|c|c|c|c|} \hline
    \multirow{2}{*}{N}                   & \multicolumn{3}{|c|}{Complete Graph}        & \multicolumn{3}{|c|}{Path}            & \multicolumn{3}{|c|}{Cycle}     \\ \cline{2-10}
    {}                    & {Constant $\Theta$}     & {Variable $\Theta$} &{Ratio}  & {Constant $\Theta$}     & {Variable $\Theta$} &{Ratio}  & {Constant $\Theta$}     & {Variable $\Theta$}  &{Ratio}   \\ \hline
    {5}  &{$1.7262$} &{$2.2616$} &{$1.3102$}   &{$0.3519$} &{$0.5173$} &{$1.4700$}   &{$0.9263$} &{$1.3053$} &{$1.4092$} \\ \hline
    {6}  &{$1.8213$} &{$2.3639$} &{$1.2979$}   &{$0.2605$} &{$0.3849$} &{$1.4775$}   &{$0.7402$} &{$1.0585$} &{$1.4300$} \\ \hline
    {7}  &{$1.8951$} &{$2.4419$} &{$1.2885$}   &{$0.1998$} &{$0.2965$} &{$1.4840$}   &{$0.5969$} &{$0.8627$} &{$1.4453$} \\ \hline
    {8}  &{$1.9539$} &{$2.5027$} &{$1.2809$}   &{$0.1578$} &{$0.2348$} &{$1.4880$}   &{$0.4874$} &{$0.7100$} &{$1.4567$} \\ \hline
    {9}  &{$2.0018$} &{$2.5519$} &{$1.2748$}   &{$0.1276$} &{$0.1901$} &{$1.4898$}   &{$0.4033$} &{$0.5907$} &{$1.4647$} \\ \hline
    {10} &{$2.0417$} &{$2.5923$} &{$1.2697$}   &{$0.1052$} &{$0.1571$} &{$1.4933$}   &{$0.3379$} &{$0.4971$} &{$1.4711$} \\ \hline
    {11} &{$2.0753$} &{$2.6257$} &{$1.2652$}   &{$0.0882$} &{$0.1318$} &{$1.4943$}   &{$0.2866$} &{$0.4230$} &{$1.4759$} \\ \hline
    {12} &{$2.1040$} &{$2.6547$} &{$1.2617$}   &{$0.0750$} &{$0.1121$} &{$1.4947$}   &{$0.2456$} &{$0.3633$} &{$1.4792$} \\ \hline
    {13} &{$2.1288$} &{$2.6790$} &{$1.2585$}   &{$0.0645$} &{$0.0965$} &{$1.4961$}   &{$0.2126$} &{$0.3151$} &{$1.4821$} \\ \hline
    {14} &{$2.1504$} &{$2.7005$} &{$1.2558$}   &{$0.0561$} &{$0.0838$} &{$1.4938$}   &{$0.1857$} &{$0.2756$} &{$1.4841$} \\  \hline
    \end{tabular}
\end{table}

\subsubsection{Path}
In a path graph with $N$ vertices, there are $N-1$ edges.
For a path topology with even number of vertices, the optimal weights are $w_{0} = 3D_L N / (2(N^2 - 1))$ and $w_{j} = ( 3D_L ( N^2  - 4j^2 ) ) / ( 2N ( N^2-1 ) )$ for $j=1, \ldots, (N/2) - 1$.
$w_{0}$ and $w_{j}$ are the weight on the middle edge and the edge that is $j$ hops away from the middle of the graph.
For a path graph with odd number of vertices, the optimal weights are $w_j  =  ( 3D_L ( N^2 - ( 2j - 1 )^2 ) ) / ( 2N ( N^2 - 1 ) )$.
$w_{j}$ is the edge that is $j$ hops away from the central vertex in the graph.
The optimal value of $\lambda_{2}(\boldsymbol{L}_{w})$ for a path graph (independent from the number of vertices) is equal to $12 D_L / \left( N \left( N^2 - 1 \right) \right)$, where $D_L$ is the upper limit on sum of the weights on the edges of the lattice core.
The optimal value of $\mu_{2,1}$ for different values of $N$ is provided in Tables 
\ref{tab:NumericalResultsDVertices} and \ref{tab:NumericalResultsDEdges}.

\subsubsection{Cycle}
For a cycle topology with $N$ vertices and $N$ edges, the optimal weight on all edges is $D_L/N$ and the optimal value of $\lambda_{2}(\boldsymbol{L}_{w})$ is equal to $2 D_L \left( 1 - cos \left( 2\pi / N \right) \right) / N$.
The optimal value of $\mu_{2,1}$ for different values of $N$ is provided in Tables 
\ref{tab:NumericalResultsDVertices} and \ref{tab:NumericalResultsDEdges}.


The results presented in Tables \ref{tab:NumericalResultsDVertices} and \ref{tab:NumericalResultsDEdges} are obtained for two different settings.
The results in Table \ref{tab:NumericalResultsDVertices}, are obtained for $D_L$ (the upper limit on sum of the weights on edges of the lattice core) set equal to the number vertices in the lattice graph $\mathcal{G}$, while the results in Table \ref{tab:NumericalResultsDEdges} are obtained for $D_L$ set equal to the number of edges in the lattice core $\mathcal{G}$.
In the case of Cycle graph, the results in Tables \ref{tab:NumericalResultsDVertices} and \ref{tab:NumericalResultsDEdges} are identical since in this topology, the number of vertices is equal to the number of edges.
For Path topology, the results are very close, since the number of edges is only one less than the number of vertices.
The most noticeable difference between the results presented in Tables \ref{tab:NumericalResultsDVertices} and \ref{tab:NumericalResultsDEdges} are those obtained for complete graph topology.
For this topology, the results obtained in Table \ref{tab:NumericalResultsDVertices} (for $D_L = N$) are decreasing as the network size is increasing,
while in the case of the results obtained in Table \ref{tab:NumericalResultsDEdges} (for $D_L = N(N-1)/2$), this trend is reversed and the convergence rate of the consensus algorithm is increasing by the size of the network.
This is due to the fact that in Table \ref{tab:NumericalResultsDVertices}, the value of $D_L$ is a first order polynomial of $N$ while in Table \ref{tab:NumericalResultsDEdges}, it is a second order polynomial of $N$.
Therefore, in the case of Table \ref{tab:NumericalResultsDEdges}, $D_L$ grows faster with the size of the network which results in faster convergence rate.
In general, the convergence rate of the consensus algorithm for the complete graph topology is much faster than that of the path and the cycle topologies.
This is due to the complete connectivity of the complete graph topology.

\section{Variable Diffusion Parameter}
\label{sec:VariableDiffusionParameter}

In previous section, the diffusion equation model of the the continuous-time consensus algorithm in the continuum limit (\ref{eq:ContinuousConsensusStateEvolution281}) is adopted and using the optimal weights obtained for the continuous-time consensus algorithm in \cite{SaberQConsensusContinuous}, the optimal weights for the diffusion equation model of the algorithm with constant diffusion parameter is derived.
%
%
In this section, we make a more comprehensive assumption than that of section \ref{sec:ConstantDiffusionParameter},
and we assume that the diffusion parameter $(\Theta)$ is not constant and it varies in terms of the spatial variable $\xi$.
This assumption has been made in an effort to achieve faster convergence rates compared to those obtained in section \ref{sec:ConstantDiffusionParameter} for constant diffusion parameter.
%
%
%
%
%
%
%
%
%
We define the variable diffusion parameter $(\Theta(\xi))$ as below,
\begin{equation}
    \label{eq:VariableThetaDefinition1092}
    \begin{gathered}
         \Theta(\xi) = \frac{3}{2} \cdot \hat{\Theta} \cdot (1-\xi^2) \quad \text{for} \quad \xi \in [0,1].
    \end{gathered}
\end{equation}
Note that the diffusion parameter $\Theta(\xi)$ defined in (\ref{eq:VariableThetaDefinition1092}) has the spatial-average value equal to the constant diffusion parameter $(\hat{\Theta})$ employed in section \ref{sec:ConstantDiffusionParameter}.
This constraint has been implemented in an effort to have a reasonable comparison between the convergence rates obtained from constant and variable diffusion parameters.
%
%
Employing the variable diffusion parameter defined in (\ref{eq:VariableThetaDefinition1092}), we obtain the following equations for the wights on the edges of the lattice $(\mathcal{W}_{\alpha\beta})$ and the weights on the edges of the path bars $(\boldsymbol{W}_{j})$,
%
%
%
%
%
\begin{subequations}
    \label{eq:WeightsContinuumConstantTheta311Variable}
    \begin{gather}
        \boldsymbol{W}_{j} = \frac{3}{2} \cdot q^2 \cdot \hat{\Theta} \cdot (1-\frac{j^2}{q^2}) \quad \text{for}\quad j=1,\cdots, q  \label{eq:WeightsContinuumConstantTheta311VariableA} \\
        \boldsymbol{W}_{\alpha\beta} = \frac{3}{2} \cdot q \cdot \hat{\Theta} \cdot \mathcal{W}_{\alpha\beta} \quad \text{for} \quad \{\alpha,\beta\}\in \mathcal{E(G)}  
    \end{gather}
\end{subequations}
%
%
%
\begin{remark}
\label{remarkWeights}
%
Interestingly, in the continuum limit (i.e. $q \rightarrow \infty$), sum of weights for both cases of constant and variable diffusion parameter case are of order $q^3$.
Based on (\ref{eq:WeightsContinuumConstantTheta311}) for sum of the weights in the case of the diffusion equation with constant diffusion parameter we have,
\begin{equation}
    \label{eq:SumWeightsConstantParameter1148}
     \begin{aligned}
        \text{Sum}_{\text{Const}}  =  |\mathcal{G}|  \sum_{j=1}^{q} \boldsymbol{W}_{j} + \sum_{\{\alpha\beta\}\in \mathcal{E}}\boldsymbol{W}_{\alpha\beta}  = 
        |\mathcal{G}|  \sum_{j=1}^{q} q^2  \hat{\Theta} + q  \hat{\Theta}  \sum_{\{\alpha\beta\}\in \mathcal{E}}\mathcal{W}_{\alpha\beta} = |\mathcal{G}|  q^3  \hat{\Theta} + o(q^3),
    \end{aligned}
\end{equation}
and sum of the weights used in the case of the diffusion equation with variable diffusion parameter (\ref{eq:WeightsContinuumConstantTheta311Variable}) are as below,
\begin{equation}
    \label{eq:SumWeightsVariableParameter1157}
    \begin{aligned}
        \text{Sum}_{\text{Var}}  &=   | \mathcal{G}|  \sum_{j=1}^{q} \boldsymbol{W}_{j} + \sum_{\{\alpha\beta\}\in \mathcal{E}}\boldsymbol{W}_{\alpha\beta} = 
        \frac{3}{2}  |\mathcal{G}|  q^2  \hat{\Theta}  \sum_{j=1}^{q} (1-\frac{j^2}{q^2}) + \frac{3}{2}  q  \hat{\Theta} \sum_{\{\alpha\beta\}\in \mathcal{E}}\mathcal{W}_{\alpha\beta} \\
        & = \frac{3}{2}  |\mathcal{G}|  q^3  \hat{\Theta}  \int_0^1 {(1-\xi^2)  d\xi} + o(q^3) = 
        |\mathcal{G}|  q^3  \hat{\Theta} + o(q^3))
    \end{aligned}
\end{equation}
\end{remark}
Using the weights in (\ref{eq:WeightsContinuumConstantTheta311Variable}), in the continuum limit, the state update equations (\ref{eq:StateUpdateEquationConsensusSpatialDiscrete265}) of the continuous-time consensus algorithm can be modelled as the following system of diffusion equations with variable diffusion parameter $\Theta(\xi)$,
\begin{subequations}
    \label{eq:StateUpdateEquationsContinuumVariableTheta1095}
    \begin{gather}
        \frac{\partial}{\partial t} Q_{\alpha}(\xi,t) = \frac{3}{2} \cdot \hat{\Theta} \frac{\partial}{\partial \xi} \left( (1-\xi^2) \cdot \frac{\partial}{\partial \xi} Q_{\alpha}(\xi,t) \right),\quad \text{for} \quad \alpha \in \mathcal{G}   \label{eq:StateUpdateEquationsContinuumVariableTheta1095a}   \\
        \frac{\partial}{\partial t} Q_{\alpha}(1,t)= - \frac{3}{2} \cdot \hat{\Theta} \cdot q \cdot (1-\xi^2) \cdot \frac{\partial}{\partial \xi} Q_{\alpha}(\xi,t)|_{\xi=1}, \quad \text{for} \quad \alpha \in \mathcal{G}         \label{eq:StateUpdateEquationsContinuumVariableTheta1095b}   \\
        \frac{\partial}{\partial t} Q_{\alpha}(0,t) = \frac{3}{2} \cdot \hat{\Theta} \cdot q \cdot {\left((1-\xi^2) \cdot   \frac{\partial}{\partial\xi}Q_{\alpha}(\xi,t)|_{\xi=0} + \sum_{\beta \in \mathcal{N}(\alpha)} \mathcal{W}_{\alpha\beta}{\left(Q_{\beta}(0,t) - Q_{\alpha}(0,t)\right)}\right)}   \label{eq:StateUpdateEquationsContinuumVariableTheta1095c}
    \end{gather}
\end{subequations}
To have finite values for $\frac{\partial}{\partial t}Q_{\alpha}(1,t)$ and $\frac{\partial}{\partial t}Q_{\alpha}(0,t)$, it is required that $(1-\xi^2) \frac{\partial}{\partial\xi}Q_{\alpha}(\xi,t)|_{\xi=1}=0$ and $\frac{\partial}{\partial\xi}Q_{\alpha}(\xi,t)|_{\xi=0}+\sum_{\beta \in \mathcal{N}(\alpha)} \mathcal{W}_{\alpha\beta} \left( Q_{\beta}(0,t) - Q_{\alpha}(0,t) \right) = 0$.
Thus the diffusion equation (\ref{eq:StateUpdateEquationsContinuumVariableTheta1095a}) and the boundary conditions (\ref{eq:StateUpdateEquationsContinuumVariableTheta1095b}) and (\ref{eq:StateUpdateEquationsContinuumVariableTheta1095c}) can be written as below,
\begin{subequations}
    \label{eq:StateUpdateEquationsContinuumVariableThetaFinalForm1351}
    \begin{gather}
        \frac{\partial}{\partial t} Q_{\alpha}(\xi,t) = \frac{3}{2} \cdot \hat{\Theta} \frac{\partial}{\partial \xi} \left( (1-\xi^2) \cdot \frac{\partial}{\partial \xi} Q_{\alpha}(\xi,t) \right),\quad \text{for} \quad \alpha \in \mathcal{G}   \label{eq:StateUpdateEquationsContinuumVariableThetaFinalForm1351a}   \\
        (1-\xi^2) \frac{\partial}{\partial\xi}Q_{\alpha}(\xi,t)|_{\xi=1} = 0, \quad \text{for} \quad \alpha \in \mathcal{G} \label{eq:StateUpdateEquationsContinuumVariableThetaFinalForm1351b}   \\
        \frac{\partial}{\partial\xi}Q_{\alpha}(\xi,t)|_{\xi=0}+\sum_{\beta \in \mathcal{N}(\alpha)} \mathcal{W}_{\alpha\beta} \left( Q_{\beta}(0,t) - Q_{\alpha}(0,t) \right) = 0, \quad \text{for} \quad \alpha \in \mathcal{G} \label{eq:StateUpdateEquationsContinuumVariableThetaFinalForm1351c}
    \end{gather}
\end{subequations}
Based on (\ref{eq:StateUpdateEquationsContinuumVariableThetaFinalForm1351}) and employing $\boldsymbol{Q}(\xi,t)$ as the vector representing the state of the diffusion system, it can be concluded that the dynamics of the diffusion system evolve according to the following diffusion equation
\begin{equation}
    \label{eq:StateUpdateEquationsContinuumVariableThetaFinalForm1117}
    \begin{gathered}
        \frac{\partial}{\partial t} \boldsymbol{Q}(\xi,t) = \frac{3}{2} \cdot \hat{\Theta} \cdot \frac{\partial}{\partial \xi} \left( (1-\xi^2) \cdot \frac{\partial}{\partial \xi} \boldsymbol{Q}(\xi,t) \right),
    \end{gathered}
\end{equation}
%
with the following boundary conditions,
\begin{subequations}
    \label{eq:StateUpdateEquationsContinuumVariableThetaFinalForm1170}
    \begin{gather}
        (1-\xi^2) \cdot \frac{\partial}{\partial \xi} \boldsymbol{Q}(\xi,t)|_{\xi = 1} = 0, \label{eq:StateUpdateEquationsContinuumVariableThetaFinalForm1170A} \\
        \frac{\partial}{\partial \xi} \boldsymbol{Q}(\xi,t)|_{\xi = 0}  =  \boldsymbol{L}_{w} \times \boldsymbol{Q}(0,t), \label{eq:StateUpdateEquationsContinuumVariableThetaFinalForm1170B}
    \end{gather}
\end{subequations}
Following similar analysis presented in section \ref{sec:ConstantDiffusionParameter}, we can state the following lemma regarding the state vector $\boldsymbol{Q}(\xi,t)$ of the diffusion system with variable diffusion parameter.

%
%
%
%
%
\begin{lemma}
\label{lemma2}
For a given weighted Laplacian matrix $\left( \boldsymbol{L}_{w} \right)$ with eigenvalues sorted as in (\ref{eq:LaplacianEigenvalues327}), the state vector $\boldsymbol{Q}(\xi,t)$ of the diffusion system (\ref{eq:StateUpdateEquationsContinuumVariableThetaFinalForm1117}) with variable diffusion parameter defined in (\ref{eq:VariableThetaDefinition1092}) can be written as below,
\begin{equation}
    \label{eq:FinalResultVariableTheta1236}
    \begin{aligned}
        \boldsymbol{Q}(\xi,t)
        =
        &\left(
        \left( \frac{1}{\sqrt{N}} \cdot \int_{0}^{1} { \boldsymbol{1}^{T} \times \boldsymbol{Q}_{0} (\xi^{'}) d\xi^{'}  } \right)   + 
        \sum_{n_{1}=0}^{\infty} A_{1,2n_1}  \cdot  e^{- \mu_{1,n_1}^{'} \cdot t} \cdot P_{2n_1}{\left(  \xi \right)}
        \cdot\right) \frac{1}{\sqrt{N}} \boldsymbol{1} 
        + \\
        %
        &
        \sum_{k=2}^{N} {
        \left(
        \sum_{n_{k} = 1}^{\infty} {  A_{k,n_{k}} \cdot e^{ - \mu_{k,n_k}^{'} \cdot t }} \cdot P_{\nu_{k,n_k}}{\left(  \xi  \right)}
        \right)
        \cdot \boldsymbol{\eta}_{k}        }
    \end{aligned}
\end{equation}
where for the coefficients $A_{1,2n_1}$ and $A_{k,n_{k}}$ we have
\begin{subequations}
    \label{eq:ACoefficients1243}
    \begin{gather}
        A_{1,2n_1}=(4n_1+1)\int_0^1 { P_{2n_1}(\xi) \cdot \left( \boldsymbol{\eta}_{1}^{T} \times \boldsymbol{Q}(\xi,0) \right) d\xi }
        \label{eq:ACoefficients1243A} \\
        A_{k,n_{k}}  =  \left(    \frac { 1 }  { \int_{0}^{1}\left( P_{\nu_{k,n_k}}{\left(  \xi  \right)}\right)^2d\xi} \right)  \cdot  \int_{0}^{1} {  \left( \boldsymbol{\eta}_{k}^{T} \times \boldsymbol{Q}(\xi,0) \right)} \cdot P_{\nu_{k,n_k}}{\left(  \xi  \right)} \cdot d\xi,
        \label{eq:ACoefficients1243B}
    \end{gather}
\end{subequations}
(\ref{eq:ACoefficients1243A}) holds for $n_1 = 0, \ldots, \infty$ and (\ref{eq:ACoefficients1243B}) holds for $k=2,\ldots,N$ and $n_{k} = 1,\ldots,\infty$.
$P_{\nu_{k,n_k}}\left(  \xi  \right)$ is the Legendre function which can be written in terms of the Hypergeometric functions as explained in Appendix \ref{sec:Hypergeometric}.
%
%
%
$\mu_{1,n_1}^{'}$ and $\mu_{k,n_k}^{'}$ are defined as below,
\begin{subequations}
    \label{eq:MuCoefficientsVariableTheta1283}
    \begin{gather}
        \mu_{1,n_1}^{'}  =  \frac{3\hat{\Theta}}{2}2n_1(2n_1+1)
        \quad \text{for} \quad n_1 = 0, \ldots, \infty  \label{eq:MuCoefficients1283A} \\
        \mu_{k,n_k}^{'}  =  \frac{3\hat{\Theta}}{2}\nu_{k,n_k}(\nu_{k,n_k}+1)
        \quad \text{for} \quad k=2,\ldots,N, \quad n_{k} = 1,\ldots,\infty  \label{eq:MuCoefficients1283B}
    \end{gather}
\end{subequations}
where the parameter $\nu_{k,n_{k}}$ for $k=2,\ldots,N$ and $n_{k} = 1,\ldots,\infty$ is obtained from the roots of the following equation,
\begin{equation}
    \label{eq:nuPolynomial1259}
    \begin{gathered}
        \frac{\nu_{k,n_{k}}(\nu_{k,n_{k}}+1)}{2}   \left(   {}_2F_1 \left( -\nu_{k,n_{k}}+1, \nu_{k,n_{k}}+2, 2, \frac{1}{2}\right) \right) =  \lambda_{k}  \left(  {}_2F_1 \left( -\nu_{k,n_{k}}, \nu_{k,n_{k}}+1, 1, \frac{1}{2} \right) \right),
    \end{gathered}
\end{equation}
\end{lemma}
The function ${}_2F_1 ( \cdot )$ is the Hypergeometric function described in appendix \ref{sec:Hypergeometric}.
Proof of this lemma is provided in Appendix \ref{sec:Proof2}.
\begin{remark}
\label{remark2}
    Based on (\ref{eq:FinalResultVariableTheta1236}) it can be concluded that the convergence rate of the state vector $\boldsymbol{Q}(\xi,t)$ to its consensus equilibrium point (\ref{eq:ConsensusPoint260}) is governed by the sentence that includes the smallest positive of $\mu_{k,n_{k}}^{'}$.
    Similar to the case with constant diffusion parameter, the parameters $\mu_{k,n_{k}}^{'}$ defined in (\ref{eq:MuCoefficientsVariableTheta1283}) have real nonnegative values since the underlying graph of the network is an undirected graph.
    $\mu_{2,1}^{'}$ is the smallest of $\mu_{k,n_{k}}^{'}$ which is positive.
    Therefore, the convergence rate of the state vector $\boldsymbol{Q}(\xi,t)$ to its consensus equilibrium point (\ref{eq:ConsensusPoint260}) is governed by $\mu_{2,1}^{'}$, defined as below,
    \begin{equation}
        \label{eq:Mu21VariableTheta1317}
        \begin{gathered}
            \mu_{2,1}^{'}  =  \frac{3}{2} \cdot \hat{\Theta} \cdot \nu_{2,1} \cdot \left( \nu_{2,1} + 1 \right).
        \end{gathered}
    \end{equation}
    where $\nu_{2,1}$ is the smallest roots of (\ref{eq:nuPolynomial1259}) for $k=2$.
\end{remark}

\begin{figure}
  \centering
     \includegraphics[width=100mm]{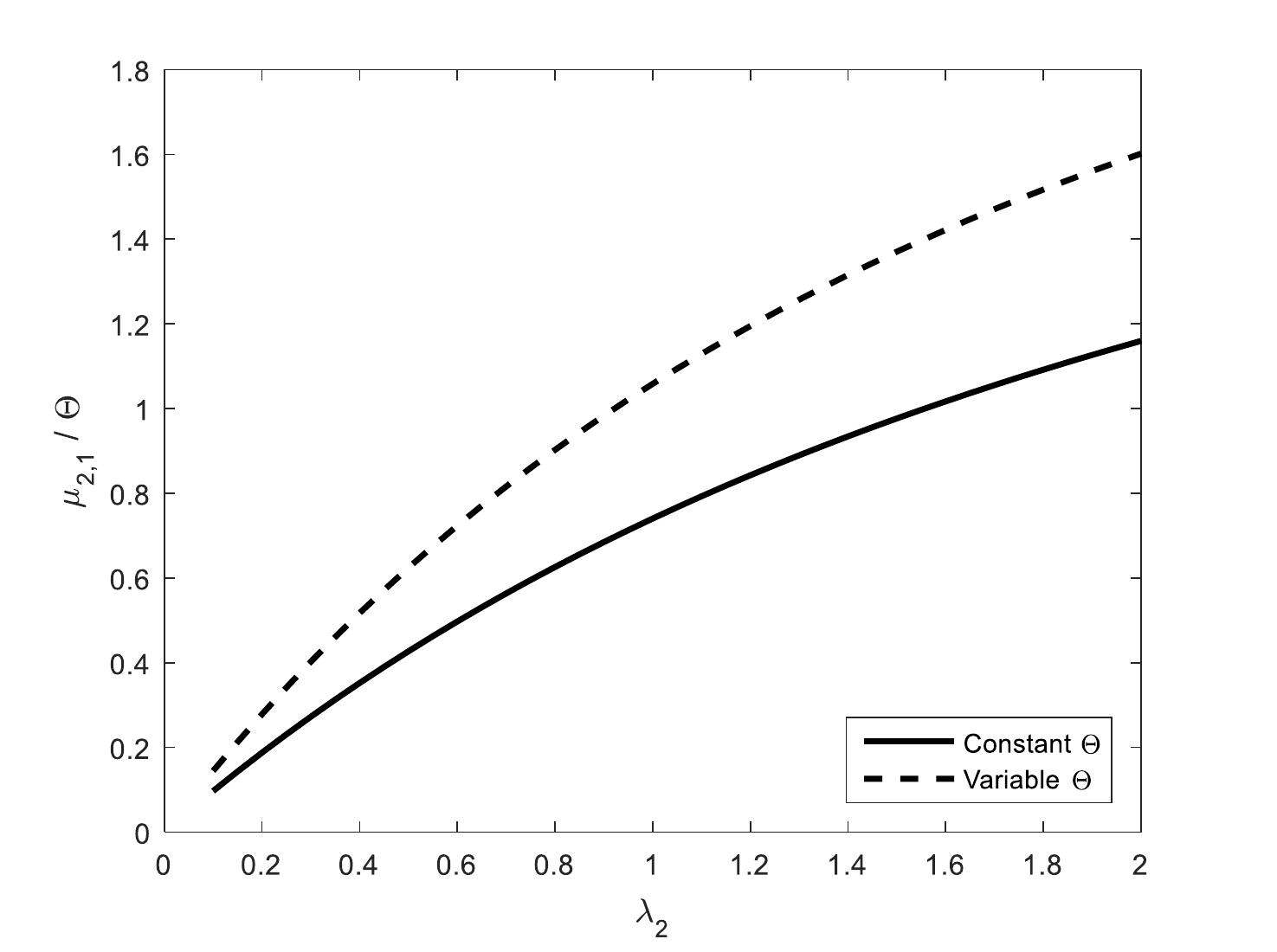}
  \caption{Values of $\mu_{2,1}/\hat{\Theta}$ (obtained from (\ref{eq:KthMuAnswerUniform558}) for constant diffusion parameter ) and $\mu_{2,1}^{'} / \hat{\Theta}$ (obtained from (\ref{eq:Mu21VariableTheta1317}) for variable diffusion parameter) in terms of $\lambda_2$ (the second smallest eigenvalue of the weighted Laplacian matrix).}
  \label{fig:Figure2MuThetaRatioCoonstVarTheta}
\end{figure}
In figure \ref{fig:Figure2MuThetaRatioCoonstVarTheta}, the values of $\mu_{2,1}/\hat{\Theta}$ (obtained from (\ref{eq:KthMuAnswerUniform558}) for constant diffusion parameter ) and $\mu_{2,1}^{'} / \hat{\Theta}$ (obtained from (\ref{eq:Mu21VariableTheta1317}) for variable diffusion parameter) in terms of $\lambda_2$ (the second smallest eigenvalue of the weighted Laplacian matrix of the lattice core) is depicted.
It is obvious from figure \ref{fig:Figure2MuThetaRatioCoonstVarTheta} that for the partially-variable diffusion parameter, significant gains are achieved in terms of the convergence rate of the diffusion system. 

\section{Symmetric Star Topology}
\label{SymmetricStarTopology}
%
%
In this section, we analyse the diffusion equations with both constant and variable diffusion parameter obtained from continuum limit of the continuous-time consensus algorithm over symmetric star topology.
%
%
%

Consider symmetric star topology with $p$ branches, each with $q$ edges. 
This topology has $|V|=1+pq$ vertices in total.
We denote agents or vertices on branches by $(\alpha, j),\quad  \alpha=1,2,\ldots,p$, $j=1,2, \ldots, q$  and the central vertex by $(0)$.
We use the notation $\boldsymbol{X}_{\alpha, j}$  for the state of the agents and $\boldsymbol{W}_{\alpha,j}$ for the weights on edges, with $\alpha=1,2,\ldots,p$, $j=1,2, \ldots, q$ and  $\boldsymbol{X}_{0}$ for the state of the central agent.
The state update equations of the continuous-time consensus algorithm \cite{SaberQConsensusContinuous} can be written as below,
\begin{subequations}
    \label{eq:ContinuousConsensusStateUpdate75}
    \begin{gather}
        \begin{array}{c} {  \frac{d}{dt}\boldsymbol{X}_{0}(t) = \sum_{\alpha=1}^{p}  { \boldsymbol{W}_{\alpha,1} \left( \boldsymbol{X}_{\alpha,1}(t)  -  \boldsymbol{X}_{0}(t) \right) }  } \end{array} \\
        \begin{array}{c} { \frac{d}{dt}\boldsymbol{X}_{\alpha,j}(t)
        =
        \boldsymbol{W}_{\alpha,j} \left( \boldsymbol{X}_{\alpha,j-1}(t)  -  \boldsymbol{X}_{\alpha,j}(t) \right)
        +
        \boldsymbol{W}_{\alpha,j+1} \left(  \boldsymbol{X}_{\alpha,j+1}(t)  -  \boldsymbol{X}_{\alpha,j}(t)  \right),
        %
        }
        \\ { j=1,2,\cdots ,q-1 } \end{array}
        \\
        \begin{array}{c} {  \frac{d}{dt}\boldsymbol{X}_{\alpha,q}(t)   =   \boldsymbol{W}_{\alpha,q} \left( \boldsymbol{X}_{\alpha,q-1}(t)  -  \boldsymbol{X}_{\alpha,q}(t)  \right).   } \end{array}
    \end{gather}
\end{subequations}
In \cite{SaberQConsensusContinuous}, the optimal weights of the continuous-time consensus algorithm are provided as below,
\begin{equation}
    \label{eq:OptimalWeightsSpatialDiscrete1588}
    \begin{gathered}
        \boldsymbol{W}_{\alpha,j}  =  \frac{3D(q+j)(q-j+1)}{pq(q+1)(2q+1)}, \quad \text{for} \quad j=1, \ldots, q.
    \end{gathered}
\end{equation}
$D$ is the upper limit on sum of the weights on edges of the whole graph.

\subsection{Symmetric Star with variable Diffusion Parameter}
\label{sec:SymStarContinuumNonUniformWeights}

In the continuum limit of $q\rightarrow \infty$,
we denote the state of the agents by $\boldsymbol{Q}_{\alpha}(\xi,t)$, i.e. $\boldsymbol{Q}_{\alpha}(\xi,t) = \boldsymbol{X}_{\alpha,j}(t)$, $\boldsymbol{Q}(0,t) = \boldsymbol{X}_{0}(t)$ and the the optimal weights (\ref{eq:OptimalWeightsSpatialDiscrete1588}) can be approximated as below,
\begin{equation}
    \nonumber
    \begin{gathered}
        \boldsymbol{W}_{\alpha}(\xi)  =  \frac{3D(1-\xi^2)}{2pq}, \quad \text{for} \quad \xi \in [0,1].
    \end{gathered}
\end{equation}
which maintains the upper limit on summation of the weights, i.e.
\begin{equation}
    \nonumber
    \begin{gathered}
        \sum_{\alpha, j} { \boldsymbol{W}_{\alpha, j} }  =  \sum_{\alpha=1}^{p} {\int_{0}^{1}\boldsymbol{W}_{\alpha}(\xi)d\xi }  = p\int_{0}^{1} { \frac{3D(1-\xi^2)}{2pq} } = D.
    \end{gathered}
\end{equation}
Using this notation,
the state update equation of the continuous-time consensus algorithm (\ref{eq:ContinuousConsensusStateUpdate75})
in the continuum limit of $q \rightarrow \infty$,
can be written as below
%
%
%
%
\begin{subequations}
    \label{eq:NewStateUpdateEqContinuum130}
    \begin{gather}
        \frac{\partial}{\partial t} \boldsymbol{Q}_{\alpha}(\xi,t) = \frac{3}{2} \hat{\Theta} \cdot \frac{\partial}{\partial \xi} \left( (1-\xi^2) \cdot  \frac{\partial}{\partial \xi}\boldsymbol{Q}_{\alpha}(\xi,t) \right), \quad \text{for} \quad \alpha=1,\cdots, p  \label{eq:NewStateUpdateEqContinuum130a} \\
        \frac{\partial}{\partial t} \boldsymbol{Q}_{\alpha}(1,t) = -\frac{3}{2} \hat{\Theta} \cdot q \cdot (1-\xi^2) \cdot \frac {\partial} {\partial \xi} \boldsymbol{Q}_{\alpha}(\xi,t)|_{\xi=1}, \quad \text{for} \quad \alpha=1,\cdots, p         \label{eq:NewStateUpdateEqContinuum130b} \\
        \frac{\partial}{\partial t} \boldsymbol{Q}(0,t)=\frac{3}{2} \hat{\Theta} \cdot q \cdot (1-\xi^2) \cdot \frac{\partial}{\partial\xi}\sum _{\alpha=1}^p \boldsymbol{Q}_{\alpha}(\xi,t)|_{\xi=0}, \label{eq:NewStateUpdateEqContinuum130c}
    \end{gather}
\end{subequations}
where we have used the constraint of the upper limit on sum of the weights i.e. $\sum_{\alpha, j} { \boldsymbol{W}_{\alpha, j} } 
= D$ and have defined $\hat{\Theta} = \frac{D}{pq^3}$.
Note that to have finite values for $\frac{\partial}{\partial t} Q_{\alpha}(1,t)$ and $\frac{\partial}{\partial t}Q(0,t)$, it is required that $\frac{\partial}{\partial\xi}Q_{\alpha}(\xi,t)|_{\xi=1}=0$ and $\frac{\partial}{\partial\xi}\sum _{\alpha=1}^pQ_{\alpha}(\xi,t)|_{\xi=0}=0$, respectively.
%
%
%
%
Consider the discrete Fourier transform of $Q_{\alpha}(\xi,t)$ given as below,
\begin{equation}
    \label{eq:DiscreteFourierTransform152}
    \begin{gathered}
       \widetilde{Q}_{\beta}(\xi,t) = \frac{1}{p}\sum _{\alpha=1}^p \omega^{\alpha\beta}Q_{\alpha}(q,t), \quad \text{for} \quad \beta=0,1,\cdots, p-1, \\
    \end{gathered}
\end{equation}
with the inverse discrete Fourier transform as below,
\begin{equation}
    \label{eq:InverseDiscreteFourierTransform160}
    \begin{gathered}
       Q_{\alpha}(\xi,t)=\sum _{\beta=0}^{p-1} \omega^{-\alpha\beta}\widetilde{Q}_{\beta}(q,t),\quad \alpha=1,2,\cdots, p \\
    \end{gathered}
\end{equation}
where $\omega=e^{-\frac{2i\pi}{p}}$.
%
%
%
%
%
%
%
%
It is straightforward to see that in terms of the $\widetilde{Q}_{\beta}(\xi,t)$, the state update equation (\ref{eq:NewStateUpdateEqContinuum130a}) and the boundary conditions (\ref{eq:NewStateUpdateEqContinuum130b}) and (\ref{eq:NewStateUpdateEqContinuum130c}) are transformed to the following diffusion equation and boundary conditions,
where for $\beta = 0$, we have
%
\begin{subequations}
    \label{eq:StateUpdateAndBoundaryCentralNode163}
    \begin{gather}
    \frac{\partial}{\partial t}\widetilde{Q}_{0}(\xi,t)=\frac{3}{2} \hat{\Theta} \cdot \frac{\partial}{\partial \xi} \left( (1-\xi^2)  \cdot  \frac{\partial}{\partial \xi}\widetilde{Q}_{0}(\xi,t)\right), 
    \label{eq:StateUpdateAndBoundaryCentralNode163a} \\
    \frac{\partial}{\partial\xi}\widetilde{Q}_{0}(\xi,t)|_{\xi=0}=0  , \quad (1-\xi^2)\frac{\partial}{\partial\xi}\widetilde{Q}_{0}(\xi,t)|_{\xi=1}=0,\quad Lim_{\xi \rightarrow 0} \widetilde{Q}_{0}(\xi,t)=pQ(0,t)  \label{eq:StateUpdateAndBoundaryCentralNode163b}
    \end{gather}
\end{subequations}
and
\begin{subequations}
    \label{eq:StateUpdateAndBoundary154}
    \begin{gather}
        \frac{\partial}{\partial t}\widetilde{Q}_{\beta}(\xi,t)=\frac{3}{2} \hat{\Theta} \cdot \frac{\partial}{\partial \xi} \left( (1-\xi^2)  \cdot  \frac{\partial}{\partial \xi}\widetilde{Q}_{\beta}(\xi,t) \right), \quad \text{for} \quad \beta=1,2,\cdots, p-1  \label{eq:StateUpdateAndBoundary154a} \\
    \widetilde{Q}_{\beta}(\xi,t)|_{\xi=0}=0  , \quad (1-\xi^2)\frac{\partial}{\partial\xi}\widetilde{Q}_{\beta}(\xi,t)|_{\xi=1}=0, \quad \text{for} \quad \beta=1,\cdots, p-1  \label{eq:StateUpdateAndBoundary154b}
    \end{gather}
\end{subequations}
%
%
%
%
%
%
By assuming that $\widetilde{Q}_{\beta}(\xi,t)$ is of the form $\widetilde{Q}_{\beta}(\xi,t)  =  e^{-\mu t} \widehat{Q}_{\beta}(\xi)$, the state update equations (\ref{eq:StateUpdateAndBoundaryCentralNode163a}) and (\ref{eq:StateUpdateAndBoundary154a}) are transformed into Legendre's differential equation as below,
\begin{equation}
    \label{eq:LegendreDifferentialEquationFormat178}
    \begin{gathered}
        \frac{d}{d\xi} \left(    (1-\xi^2) \frac{d}{d\xi} \widehat{Q}_{\beta}(\xi)  \right)  +  \frac{2\mu}{3\hat{\Theta}}  \widehat{Q}_{\beta}(\xi)    =  0. \quad \text{for} \quad  \beta = 0,1,\ldots,p-1.
    \end{gathered}
\end{equation}
As explained in appendix \ref{sec:Legendre},
if $\frac{2\mu_n}{3\hat{\Theta}}  =  n(n+1)$ where $n$ is a non-negative integer, the answer to equation (\ref{eq:LegendreDifferentialEquationFormat178}) is of the polynomial form, i.e. $\widehat{Q}_{\beta}(\xi) = P_{n}(\xi)$, where $P_{n}(\xi)$ is the Legendre polynomial of order $n$.
%
%
Note that the reason we consider only the polynomial solutions of the Legendre differential equations (\ref{eq:LegendreDifferentialEquationFormat178}) is that for the polynomial solution, $\lim_{\xi \rightarrow 1}(1-\xi^2)\frac{d}{d\xi}Q_{\alpha}(\xi)$ in the right hand side of (\ref{eq:NewStateUpdateEqContinuum130b}) is zero.
For non-polynomial solution this limit has a finite value resulting in a infinite value for the right hand side of (\ref{eq:NewStateUpdateEqContinuum130b}) (since in the continuum limit, $q \rightarrow \infty$) which is not acceptable.
%
%
%
%
Thus the spectrum of (\ref{eq:StateUpdateAndBoundaryCentralNode163a}) and (\ref{eq:StateUpdateAndBoundary154a}) are as below,
\begin{subequations}
    \label{eq:EigenvalueNonUniform123}
    \begin{gather}
       \mu_{2k}=\frac{3 \hat{\Theta} }{2}2k(2k+1), \quad k=0,1,2,\cdots, \label{eq:EigenvalueNonUniform123A}\\
       \mu_{2k+1}=\frac{3 \hat{\Theta} }{2}2(k+1)(2k+1), \quad k=0,1,2,\cdots,\quad \text{with degeneracy} \; p-1. \label{eq:EigenvalueNonUniform123B}
    \end{gather}
\end{subequations}
%
%
%
%
%
%
and the solution to the state update equations (\ref{eq:StateUpdateAndBoundaryCentralNode163a}) and (\ref{eq:StateUpdateAndBoundary154a}) are as below,
\begin{subequations}
    \label{eq:SolutionStateUpdateEquations204}
    \begin{gather}
       \widetilde{Q}_0(\xi,t)=\sum_{k=0}^{\infty} \widetilde{A}_{2k}(0) \cdot e^{-3k(2k+1)\hat{\Theta} \cdot t} \cdot P_{2k} (\xi)        \\
       \widetilde{Q}_{\beta}(\xi,t) = \sum_{k=0}^{\infty}  { \widetilde{A}_{2k+1}(\beta) \cdot e^{-3(k+1)(2k+1)\hat{\Theta} \cdot t} \cdot P_{2k+1} (\xi)  },  \quad  \text{for}  \quad  \beta=1,\cdots, p-1
    \end{gather}
\end{subequations}
%
%
with the coefficients $\widetilde{A}_{2k}(0)$ and $\widetilde{A}_{2k+1}(\beta)$ defined as below,
\begin{subequations}
    \label{eq:ACoefficients225}
    \begin{gather}
       \widetilde{A}_{2k}(0) = (4k+1) \cdot \int_{0}^{1} { P_{2k}(\xi) \cdot \widetilde{Q}_{0}(\xi,0) \cdot d\xi }        \\
       \widetilde{A}_{2k+1}(\beta) = 2(2k+1) \cdot \int_{0}^{1} { P_{2k+1}(\xi) \cdot \widetilde{Q}_{\beta}(\xi,0) \cdot d\xi }, \quad \text{for} \quad \beta=1,\cdots, p-1,
    \end{gather}
\end{subequations}
Substituting the coefficients $\widetilde{A}_{2k+1}(\beta)$ and $\widetilde{A}_{2k}(0)$ from (\ref{eq:ACoefficients225}) in (\ref{eq:SolutionStateUpdateEquations204}) and the resultant in (\ref{eq:InverseDiscreteFourierTransform160}), we obtain the following as the final answer for the state variable $Q_{\alpha}(\xi,t)$ 
with variable diffusion parameter,
\begin{equation}
    \label{eq:OrthogonalRelations373_2047}
 \begin{gathered}
    \resizebox{.90\hsize}{!}{$Q_{\alpha}(\xi,t)
    =
    \frac{1}{p}\int_0^1 { \sum _{\beta=1}^{p} {Q_{\beta}(\xi^{\prime},0) } \cdot d\xi^{\prime} }  
    +
    \frac{1}{p}\sum_{k=1}^{\infty}(4k+1)\int_0^1 { P_{2k}(\xi^{\prime}) \cdot Q_{0}(\xi^{\prime},0) \cdot e^{-3k(2k+1)\hat{\Theta} t} \cdot P_{2k} (\xi) \cdot d\xi^{\prime} } $}
    \\
    \hspace{-20pt}
    \resizebox{.70\hsize}{!}{$ + \frac{1}{p}\sum_{k=0}^{\infty} (4k+3) \cdot \int_0^1 P_{2k}(\xi^{\prime}) \cdot \sum _{\beta=1}^{p} \left( Q_{\alpha}(\xi^{\prime},0) - Q_{\beta}(\xi^{\prime},0) \right) \cdot e^{-3(k+1)(2k+1) \cdot \hat{\Theta} \cdot t} \cdot P_{2k+1}(\xi^{\prime}) \cdot  d\xi^{\prime} $} 
    .
 %
%
    %
    \end{gathered}
\end{equation}
$\mu_{1} = 3 \hat{\Theta}$ governs the convergence rate of $Q_{\alpha}(\xi,t)$ to its equilibrium value.
This is due to the fact that among all exponential functions $e^{-\mu_{n} t}$, $e^{-\mu_{1} t}$ has the slowest convergence rate to zero.
%
The functionality of $\mu_{1}$ in governing the convergence rate is similar to that of the second smallest eigenvalue of the weighted Laplacian matrix in the continuous-time consensus algorithm \cite{SaberQConsensusContinuous}.
%
%
%
The second smallest eigenvalue of the weighted Laplacian matrix in the continuous-time consensus algorithm over symmetric star topology with optimal (non-uniform) weights is as below,
\begin{equation}
    \label{eq:SymmetricStarLambda2Optimal273}
    \begin{gathered}
       \lambda_2=\frac{6D}{pq(q+1)(2q+1)},
    \end{gathered}
\end{equation}
in the continuum limit that $q\rightarrow \infty$, the value of this eigenvalue can be written as below, 
\begin{equation}
    \label{eq:SymmetricStarLambda2OptimalContinuum281}
    \begin{gathered}
       \lim_{q\rightarrow \infty}\lambda_2 = \lim_{q\rightarrow \infty}\frac{6D}{pq(q+1)(2q+1)} = \frac{3D}{pq^3} = 3\hat{\Theta}.
    \end{gathered}
\end{equation}
Considering the previously chosen value of $\hat{\Theta} = \frac{D}{pq^3}$, it is obvious that the results in (\ref{eq:SymmetricStarLambda2OptimalContinuum281}) are in agreement with those obtained for the convergence rate of $Q_{\alpha}(\xi,t)$ to its equilibrium value.
%


\subsection{Symmetric Star with Constant Diffusion Parameter}
\label{sec:SymStarContinuumUniformWeights}

Considering the description of the symmetric star topology and using the notations introduced in the beginning of this section, we can write the following for the state update equations of the continuous-time consensus algorithm (\ref{eq:ContinuousConsensusStateUpdate75}),
%
%
%
%
\begin{subequations}
    \label{eq:ConsensusEquation235}
    \begin{gather}
        \frac{d}{dt}X_{0}(t) = - \eta \cdot {\left( p \cdot X_{0}(t)-\sum_{\alpha=1}^{p} X_{\alpha,1}(t)\right)}  \label{eq:ConsensusEquation235a}\\
        \frac{d}{dt}X_{\alpha,j}(t) = - \eta \cdot \left( 2X_{\alpha,j}(t) - X_{\alpha,j+1}(t) - X_{\alpha,j-1}(t) \right) 
        \label{eq:ConsensusEquation235b}\\
        \frac{d}{dt}X_{\alpha,q}(t) = - \eta \cdot \left( X_{\alpha,q}(t) - X_{\alpha,q-1}(t) \right) \label{eq:ConsensusEquation235c}
    \end{gather}
\end{subequations}
where equation (\ref{eq:ConsensusEquation235b}) is for $j=1,\ldots,q-1$ 
and $\eta$ denotes the constant weight on the edges.
Similar to subsection \ref{sec:SymStarContinuumNonUniformWeights}, in the continuum limit of $q \rightarrow \infty$ we denote the state of vertices by $Q_{\alpha}(\xi,t)$ i.e. $Q_{\alpha}(\xi,t) = X_{\alpha,j}(t)$ and $Q_{\alpha}(0,t)=X_{0}(t)$.
Thus the state update equations (\ref{eq:ConsensusEquation235})  in the continuum limit of $q \rightarrow \infty$ can be written as below,
\begin{subequations}
    \label{eq:OrthogonalRelationsContinuum251}
    \begin{gather}
        \frac{\partial}{\partial t}Q_{\alpha}(\xi,t) = \hat{\Theta} \cdot \frac{\partial^2}{\partial\xi^2}Q_{\alpha}(\xi,t), \quad \text{for} \quad \alpha=1,2,\cdots, p,  \label{eq:OrthogonalRelationsContinuum251a}  \\
        \frac{\partial}{\partial t}Q_{\alpha}(q,t)=- \hat{\Theta} \cdot q \cdot \frac{ \partial } { \partial \xi } Q_{\alpha}(\xi,t)|_{\xi=1}, \quad \text{for} \quad \alpha=1,2,\cdots, p,         \label{eq:OrthogonalRelationsContinuum251b} \\
        \frac{\partial}{\partial t}Q_{\alpha}(0,t) = \hat{\Theta} \cdot q \cdot \frac{\partial}{\partial\xi}\sum _{\alpha=1}^pQ_{\alpha}(\xi,t)|_{\xi=0},  \label{eq:OrthogonalRelationsContinuum251c}
    \end{gather}
\end{subequations}
where we have used the constraint that $D$ is the upper limit on sum of the weights 
i.e. $(1+pq)\eta \approx pq\eta = D $ and
similar to subsection \ref{sec:SymStarContinuumNonUniformWeights}, $\hat{\Theta}$ is defined as $\hat{\Theta} = \frac{D}{pq^3}$.
%
%
%
Considering the discrete Fourier transform of $Q_{\alpha}(\xi,t)$ as in (\ref{eq:DiscreteFourierTransform152}) and its inverse (\ref{eq:InverseDiscreteFourierTransform160}), it can be shown that
the state update equation (\ref{eq:OrthogonalRelationsContinuum251a}) and the boundary conditions (\ref{eq:OrthogonalRelationsContinuum251b}) and (\ref{eq:OrthogonalRelationsContinuum251c}) are transformed to the following diffusion equation and boundary conditions,
where for $\beta = 0$,
\begin{subequations}
    \label{eq:BoundaryConditionDiscreteFourierTransformCentre302}
    \begin{gather}
    \frac{\partial}{\partial t}\widetilde{Q}_{0}(\xi,t) = \hat{\Theta} \cdot \frac{\partial^2}{\partial\xi^2}\widetilde{Q}_{0}(\xi,t),  \label{eq:BoundaryConditionDiscreteFourierTransformCentre302a} \\
    \frac{\partial}{\partial\xi}\widetilde{Q}_{0}(\xi,t)|_{\xi=0}=0  , \quad \frac{\partial}{\partial\xi}\widetilde{Q}_{0}(\xi,t)|_{\xi=1}=0,\quad \lim_{\xi \rightarrow 0} \widetilde{Q}_{0}(\xi,t)=p \cdot Q(0,t)  \label{eq:BoundaryConditionDiscreteFourierTransformCentre302b}
    \end{gather}
\end{subequations}
and
\begin{subequations}
    \label{eq:BoundaryConditionDiscreteFourierTransform291}
    \begin{gather}
    \frac{\partial}{\partial t} \widetilde{Q}_{\beta}(\xi,t) = \hat{\Theta} \frac{\partial^2}{\partial\xi^2} \widetilde{Q}_{\beta}(\xi,t), \quad \text{for} \quad \beta = 1,\ldots,p       \label{eq:BoundaryConditionDiscreteFourierTransform291a}
    \\
    \widetilde{Q}_{\beta}(\xi,t)|_{\xi=0}=0, \quad 
    \frac{\partial}{\partial\xi}\widetilde{Q}_{\beta}(\xi,t)|_{\xi=1}=0. 
    \label{eq:BoundaryConditionDiscreteFourierTransform291b}
    \end{gather}
\end{subequations}
Solving (\ref{eq:BoundaryConditionDiscreteFourierTransformCentre302}) and (\ref{eq:BoundaryConditionDiscreteFourierTransform291}),
we obtain the following for the eigenvalues and the general solution of (\ref{eq:BoundaryConditionDiscreteFourierTransformCentre302}) and (\ref{eq:BoundaryConditionDiscreteFourierTransform291})
\begin{subequations}
    \label{eq:Eigenvalues68}
    \begin{gather}
       \mu_{2k}=\frac{(2k\pi)^2\hat{\Theta}}{4}, \quad k=1,2,\cdots, \\
       \mu_{2k+1}=\frac{((2k+1)\pi)^2\hat{\Theta}}{4}, \quad k=0,1,2,\cdots,\quad \text{with degeneracy} \; p-1
    \end{gather}
\end{subequations}
%
%
%
and
\begin{subequations}
    \label{eq:AnswerDiscreteFourierTransformUniformWeight471}
 \begin{gather}
       \widetilde{Q}_0(\xi,t)
       =
       \sum_{k=0}^{\infty} \widetilde{A}_{2k}(0) \cdot e^{-\frac{(2k\pi)^2 \cdot \hat{\Theta} \cdot t}{4}} \cdot \cos(\frac{2k\pi \xi}{2})
        \label{eq:AnswerDiscreteFourierTransformUniformWeight471A}
       \\
       \widetilde{Q}_{\beta}(\xi,t)
       =
       \sum_{k=0}^{\infty}\widetilde{A}_{2k+1}(\beta) \cdot e^{-\frac{((2k+1)\pi)^2 \cdot \hat{\Theta} \cdot t}{4}} \cdot \sin{(\frac{(2k+1) \cdot \pi \cdot \xi}{2})},\quad \text{for} \quad \beta=1,2,\cdots, p-1,   \label{eq:AnswerDiscreteFourierTransformUniformWeight471B}
    \end{gather}
\end{subequations}
with the coefficients $\widetilde{A}_{2k+1}(\beta)$ and $\widetilde{A}_{2k}(0)$ as below,
\begin{subequations}
    \label{eq:ACoefficientsUniformWeights491}
    \begin{gather}
       \widetilde{A}_{0}(0)=\int_0^1 { \widetilde{Q}_{0}(\xi,0) \cdot d\xi }
       \\
       \widetilde{A}_{2k}(0)=\frac{1}{2}\int_0^1 \cos{(\frac{2k\pi \xi}{2})} \cdot \widetilde{Q}_{0}(\xi,0) \cdot d\xi, \quad \text{for} \quad k\neq 0,
       \\
       \widetilde{A}_{2k+1}(\beta) = \frac{1}{2} \int_0^1 { \sin{\left( \frac{(2k+1) \cdot \pi \cdot \xi} {2} \right) } \cdot \widetilde{Q}_{\beta}(\xi,0) \cdot d\xi }, \quad \text{for} \quad \beta=1,2,\cdots, p-1
    \end{gather}
\end{subequations}
Substituting the coefficients $\widetilde{A}_{2k+1}(\beta)$ and $\widetilde{A}_{2k}(0)$ from (\ref{eq:ACoefficientsUniformWeights491}) in 
(\ref{eq:AnswerDiscreteFourierTransformUniformWeight471}) and the resultant in (\ref{eq:InverseDiscreteFourierTransform160}) we obtain the following as the final answer for the state variable $Q_{\alpha}(\xi,t)$ 
with constant diffusion parameter,
%
\begin{equation}
    \label{eq:FinalAnswerUniformWeights505}
 \begin{gathered}
      \resizebox{.90\hsize}{!}{$
      Q_{\alpha}(\xi,t)       =  
      \frac{1}{p}\int_0^1 { \sum _{\beta=1}^p Q_{\beta}(\xi^{\prime},0) \cdot d\xi^{\prime} } 
      +
      \frac{1}{2p}\sum_{k=1}^{\infty} { \left( \int_0^1 \cos{(\frac{2k\pi \xi^{\prime}}{2}) } \cdot \sum _{\beta=1}^p Q_{\beta}(\xi^{\prime},0)\cdot d\xi^{\prime} \right) \cdot e^{-\frac{(2k\pi)^2 \cdot \hat{\Theta} \cdot t}{4}} \cdot \cos{(\frac{2k\pi \xi}{2})} }
      $} 
      \\ 
      \hspace{-20pt}
      \resizebox{.70\hsize}{!}{$+ \frac{1}{2p}\sum_{k=0}^{\infty} \left( \int_0^1 \sin{(\frac{(2k+1)\pi \xi^{\prime}}{2})} \cdot \sum _{\beta=1}^p \left( Q_{\alpha}(\xi^{\prime},0) - Q_{\beta}(\xi^{\prime},0) \right) \cdot d\xi^{\prime} \right) \cdot e^{-\frac{((2k+1)\pi)^2 \cdot \hat{\Theta} \cdot t}{4}} \cdot \sin{(\frac{(2k+1)\pi \xi}{2})}  $}
    \end{gathered}
\end{equation}

\subsection{Optimizing the Diffusion Parameter by Variational Method}
\label{sec:VariationalProof}



In this subsection, we try to obtain the optimal 
diffusion parameter $\Theta(\xi)$ for a network with
symmetric star topology using variational method. 
We consider the symmetric star topology introduced in the beginning of this section. 
For the given diffusion parameter $\Theta(\xi)$,
the state update equation of the continuous-time consensus algorithm with symmetric star topology (\ref{eq:ContinuousConsensusStateUpdate75}) in the continuum limit can be written as the following
diffusion equation with Von-Neumann boundary conditions, 
\begin{subequations}
    \label{eq:DiffusionPDEStateUpdateTimeVariational73}
    \begin{gather}
        \frac{\partial}{\partial t}Q_{\alpha}(\xi,t)=  \frac{\partial}{\partial \xi} \left(\Theta(\xi)  \frac{\partial}{\partial \xi}Q_{\alpha}(\xi,t) \right), \quad \text{for} \quad \alpha=1,2,\cdots, p  \\
       \Theta(\xi)\frac{\partial}{\partial\xi}Q_{\alpha}(\xi,t)|_{\xi=1}=0, \quad \text{for} \quad \alpha=1,2,\cdots, p         \\
       \Theta(\xi) \frac{\partial}{\partial\xi}\sum _{\alpha=1}^pQ_{\alpha}(\xi,t)|_{\xi=0}=0
    \end{gather}
\end{subequations}
%
Our aim is to obtain the optimal function for the spatially-variable diffusion parameter $\Theta(\xi)$ with the constraint $\int_{0}^{1} \Theta(\xi)d\xi = \hat{\Theta}$ on its average value, that optimizes the convergence rate of the state vector $Q(\xi,t)$ to its equilibrium state (\ref{eq:ConsensusPoint260}).
To do so, first we separate the time-dependent part of $Q_{\alpha}(\xi,t)$ by setting $Q_{\alpha}(\xi,t)= e^{-\mu \cdot t} Q_{\alpha}(\xi)$, for $\alpha=1,\cdots, p$, which results in the following 
\begin{equation}
    \label{eq:DiffusionPDEStateUpdateNoTimeVariational86}
    \begin{gathered}
        \frac{d}{d \xi} \left(\Theta(\xi)  \frac{d}{d \xi}Q_{\alpha}(\xi)\right)+\mu Q_{\alpha}(\xi)=0,\quad \text{for }\quad \alpha=1,\cdots, p \\
         \Theta(\xi)\frac{d}{d\xi}Q_{\alpha}(\xi)|_{\xi=1}=0  \quad \text{for }\quad \alpha=1,\cdots, p       \\
       \Theta(\xi) \Sigma_{\alpha=1}^p\frac{d}{d \xi}Q_{\alpha}(\xi)|_{\xi=0}=0
        \end{gathered}
\end{equation}
%
To solve above differential equations, for a given diffusion parameter $\Theta(\xi)$, 
we use the discrete Fourier transform of $Q_{\alpha}\left( \xi \right)$ as provided in (\ref{eq:DiscreteFourierTransform152}) and its inverse discrete Fourier transform (\ref{eq:InverseDiscreteFourierTransform160}) to write the differential equation and the boundary conditions (\ref{eq:DiffusionPDEStateUpdateNoTimeVariational86}) in terms of $\widetilde{Q}_{\beta}(\xi)$ as below,
%
%
%
\begin{subequations}
    \label{eq:DiffusionPDEFourierTransformBetaZeroVariational132}
    \begin{gather}
        \frac{d}{d \xi} \left( \Theta(\xi)  \frac{d}{d \xi}\widetilde{Q}_{0}(\xi)\right)+\mu \widetilde{Q}_{0}(\xi)=0,  \\
    \Theta(\xi)\frac{d}{d\xi}\widetilde{Q}_{0}(\xi)|_{\xi=0}=0  , \quad \Theta(\xi)\frac{d}{d\xi}\widetilde{Q}_{0}(\xi)|_{\xi=1}=0,\quad
    Lim_{\xi \rightarrow 0} \widetilde{Q}_{0}(\xi)=Q(0)
    \end{gather}
\end{subequations}
and
\begin{subequations}
    \label{eq:DiffusionPDEFourierTransformBetaNonZeroVariational122}
    \begin{gather}
    \frac{d}{d \xi} \left(\Theta(\xi)  \frac{d}{d \xi}\widetilde{Q}_{\beta}(\xi)\right)+\mu \widetilde{Q}_{\beta}(\xi)=0, \quad \text{for} \quad \beta=1,2,\cdots, p-1\\
    \widetilde{Q}_{\beta}(\xi)|_{\xi=0}=0, \quad   \Theta(\xi)\frac{d}{d\xi}\widetilde{Q}_{\beta}(\xi)|_{\xi=1}=0,\quad \text{for} \quad \beta=1,2,\cdots, p-1
     \end{gather}
\end{subequations}
%
%
%
Solving equations (\ref{eq:DiffusionPDEFourierTransformBetaZeroVariational132}) and (\ref{eq:DiffusionPDEFourierTransformBetaNonZeroVariational122}), we obtain the spectrum of the corresponding Sturm-Liouville operator (i.e. $\frac{d}{d \xi} \Theta(\xi)  \frac{d}{d\xi}$ ) with two different von-Neumann boundary conditions as below,
%
%
\begin{subequations}
    \label{eq:DiffusionPDESturmLiouvilleBetaNonZeroVariational148}
    \begin{gather}
    \frac{d}{d \xi} \left( \Theta(\xi)  \frac{d}{d \xi}\Phi_{1,k_1}(\xi)\right)+\mu_{1,k_1} \Phi_{1,k_1}(\xi)=0,  \\
    \Theta(\xi)\frac{d}{d\xi}\Phi_{1,k_1}(\xi)|_{\xi=0}=0  , \quad \Theta(\xi)\frac{d}{d\xi}\Phi_{1,k_1}(\xi)|_{\xi=1}=0,\quad \text{for}\quad k_1=0,1,\cdots,\infty\\
    \int_0^1\Phi_{1,k_1}(\xi)\Phi_{1,k_1^{\prime}}(\xi)=\delta_{k_1,k_1^{\prime}},
    \end{gather}
\end{subequations}
and
\begin{subequations}
    \label{eq:DiffusionPDESturmLiouvilleBetaZeroVariational158}
    \begin{gather}
    \frac{d}{d \xi} \left(\Theta(\xi)  \frac{d}{d \xi}\Phi_{2,k_2}(\xi)\right)+\mu_{2,k_2} \Phi_{2,k_2}(\xi)=0\\
    \Phi_{2,k_2}(\xi)|_{\xi=0}=0, \quad   \Theta(\xi)\frac{d}{d\xi}\Phi_{2,k_2}(\xi)|_{\xi=1}=0,
     \quad \text{for}\quad k_2=0,1,\cdots,\infty \\
     \int_0^1\Phi_{2,k_2}(\xi)\Phi_{2,k_2^{\prime}}(\xi)=\delta_{k_2,k_2^{\prime}},
      \end{gather}
\end{subequations}
The Sturm-Liouville operator $\frac{d}{d \xi} \Theta(\xi)  \frac{d}{d\xi}$ with above given  boundary conditions is self adjoint, i.e. for two functions $\Psi$ and $\Psi^{\prime} $ satisfying either set of above boundary conditions we have
\begin{equation}
    \label{eq:SturmLiouvilleSelfAdjointVariational170}
    \begin{gathered}
       \int_0^1 d\xi\Psi(\xi)\frac{d}{d \xi} \left(\Theta(\xi)  \frac{d}{d \xi}\Psi^{\prime}(\xi)\right)=\int_0^1 d\xi\Psi^{\prime}(\xi)\frac{d}{d \xi} \left(\Theta(\xi)  \frac{d}{d \xi}\Psi(\xi)\right)
      \end{gathered}
\end{equation}
Therefore, its eigenvalues $ \mu_{1,k_1}$  and $ \mu_{2,k_2}$ are real and  any pair of eigenfunctions associated with distinct eigenvalues are orthogonal. 
%
The eigenfunction $ \Phi_{1,k_1}(\xi)$ corresponding to eigenvalue $\mu_{1,k_1}$ is unique (up to a normalization constant) with exactly $k_1$ zeros in the interval  $(0, 1)$.
Similarly, the eigenfunction $ \Phi_{2,k_2}(\xi)$ corresponding to eigenvalue $\mu_{2,k_2}$ is also unique (up to a normalization constant) but with exactly $k_2 - 1$ zeros in the interval  $(0, 1)$.
%
Furthermore, we have $\mu_{1,1}=0$ and $\Phi_{1,1}=1$.

From (\ref{eq:DiffusionPDESturmLiouvilleBetaNonZeroVariational148}) and (\ref{eq:DiffusionPDESturmLiouvilleBetaZeroVariational158}), it can be concluded that 
the convergence rate of the state vector $Q(\xi,t)$ to its equilibrium state (\ref{eq:ConsensusPoint260}) is governed by $\min \{\mu_{1,2}, \mu_{2,1}\}$.
%
Our aim is 
to find the 
optimal function for the diffusion parameter $\Theta(\xi)$ that maximizes 
the convergence rate of the state vector $Q(\xi,t)$ to its equilibrium state (\ref{eq:ConsensusPoint260}).
i.e., a function for the diffusion parameter $\Theta(\xi)$ that maximizes $\text{min} \{\mu_{1,2}, \mu_{2,1}\}$.
Obviously $\mu_{1,2} > \mu_{2,1}$  (see Appendix \ref{SturmLiouvilleAppendix} for proof), therefore, we only have to maximize $\mu_{2,1}$. 
%
$\mu_{2,1}$ can be obtained using variational method \cite{Simon1978,RiemannZetaRef1,LegendreBookHilbert1965}, i.e. minimizing the Rayleigh quotient  $ \int_{0}^{1} \Phi(\xi) \frac{d}{d\xi} \Theta(\xi) \frac{d}{d\xi} \Phi(\xi) d\xi   /  \int_0^1(\Phi(\xi))^2d\xi $.
This is equivalent to maximizing $\int_{0}^{1} {\Theta(\xi) \left( \frac{d\Phi(\xi)}{d\xi} \right)^2 d\xi} / \int_0^1(\Phi(\xi))^2d\xi$, since using the integration by parts method, the integral $\int_{0}^{1} {\Phi(\xi) \frac{d}{d\xi} \Theta(\xi) \frac{d}{d\xi} \Phi(\xi) d\xi}$ can be written as $-\int_{0}^{1} {\Theta(\xi) \left( \frac{d\Phi(\xi)}{d\xi} \right)^2 d\xi}$.
Therefore calculating $\mu_{2,1}$ reduces to the following optimization problem,
%
%
\begin{equation}
    \label{eq:MaximizationProblemVariational214}
    \begin{aligned}
        \max\limits_{\boldsymbol{\Theta(\xi), \Phi(\xi)}} 
        &\int_0^1\Theta(\xi)\left(\frac{d\Phi(\xi)}{d\xi}\right)^2 \cdot d\xi \\
        s.t. \quad &\int_0^1(\Phi(\xi))^2d\xi = 1, \quad \int_0^1\Theta(\xi)d\xi = \hat{\Theta},
    \end{aligned}
\end{equation}
with boundary conditions $ \Phi(0)=0$  and  $\Theta(\xi) \frac{d}{d \xi}\Phi(\xi)|_{\xi=1}=0$. 
By introducing the relevant Lagrange multipliers $\left( \mu , \Omega \right)$, all we need is to maximize the following
 \begin{equation}
    \label{eq:MaximizationProblemLagrangeVariational225}
    \begin{gathered}
       \text{maximize} \quad \int_0^1\Theta(\xi)\left(\frac{d\Phi(\xi)}{d\xi}\right)^2 +
        \mu\left(1-\int_0^1(\Phi(\xi))^2\right)+ 
        \Omega\left(\hat{\Theta}-\int_0^1\Theta(\xi)d\xi\right),
    \end{gathered}
\end{equation}
with boundary conditions $ \Phi(0)=0$  and  $\Theta(\xi) \frac{d}{d \xi}\Phi(\xi)|_{\xi=1}=0$.
To this aim, we have to set the variation of (\ref{eq:MaximizationProblemLagrangeVariational225}) (with respect to $\Phi(\xi)$ and $\Theta(\xi))$ to zero , i.e.
\begin{equation}
    \label{eq:Eq12Variational235}
    \begin{gathered}
        -2\int_0^1d\xi \delta \Phi(\xi)\left(\frac{d}{d \xi} \left(\Theta(\xi)  \frac{d}{d \xi}\Phi(\xi)\right) +\mu \Phi(\xi)\right)+
        \\
        \int_0^1d\xi \delta \Theta(\xi)\left(\left(\frac{d\Phi(\xi)}{d\xi}\right)\left(\frac{d\Phi(\xi)}{d\xi}\right)-\Omega\right) +\Theta(\xi)\delta\Phi(\xi)\frac{d}{d \xi}\Phi(\xi)|_{\xi=0}^{\xi=1}
    \end{gathered}
\end{equation}
The last term $ \Theta(\xi)\delta\Phi(\xi)\frac{d}{d \xi}\Phi(\xi)|_{\xi=0}^{\xi=1}$ is zero because $\Theta(\xi)\delta\Phi(\xi)\frac{d}{d \xi}\Phi(\xi)|_{\xi=1}=0 $ and $\delta \Phi(\xi)|_{\xi=0}$ $=0$.
The variation of Lagrange variables leads to the constraints in (\ref{eq:MaximizationProblemVariational214}).
Now due to arbitrariness of  $\delta \Phi(\xi)$ and $\delta \Theta(\xi)$ we have:
\begin{equation}
    \label{eq:Eq13Variational247}
    \begin{gathered}
        \frac{d}{d \xi} \left(\Theta(\xi)  \frac{d}{d \xi}\Phi(\xi)\right) +\mu \Phi(\xi)=0
        \\
        \left(\frac{d\Phi(\xi)}{d\xi}\right)^2=\Omega
    \end{gathered}
\end{equation}
Now substituting $ \left(\frac{d\Phi(\xi)}{d\xi}\right)^2=\Omega $ in $ \int_0^1(\Phi(\xi))^2d\xi=1$  we obtain $\Omega =3$ and consequently $\frac{d\Phi(\xi)}{d\xi}=\sqrt{3} $ and integrating it we get $\Phi_{2,1}(\xi)=\sqrt{3}\xi$.
Substituting $\Phi(\xi)=\Phi_{2,1}(\xi)=\sqrt{3}\xi$ in above equation for the diffusion parameter $\Theta(\xi)$ we obtain $\frac{d\Theta(\xi)}{d \xi}+\mu \xi=0$ and integrating it we obtain $\Theta(\xi)=-\frac{\mu}{2}\xi^2+cte$ where using
$\Theta(\xi) \frac{d}{d \xi}\Phi(\xi)|_{\xi=1}=0$ we get $cte=\frac{\mu}{2}$ and finally using $ \int_0^1\Theta(\xi)d\xi=\hat{\Theta}$ we obtain the following final results which is in agreement with those obtained by taking the continuum limit of continuous time consensus of symmetric star,
\begin{equation}
    \label{eq:Eq14Variational262}
    \begin{gathered}
       \mu=3\hat{\Theta}, \quad \Theta(\xi)=\frac{3}{2}\hat{\Theta}(1-\xi^2), \quad \Phi_{2,1}(\xi) 
       =\sqrt{3}\xi
    \end{gathered}
\end{equation}
the Lagrange coefficient $\mu$ is also the second smallest eigenvalue (i.e. $\mu_{2,1}$).

\subsection{Robustness of the Diffusion System} 
In this subsection, we investigate the robustness of the diffusion system 
over the symmetric star topology with both variable (\ref{eq:NewStateUpdateEqContinuum130}) and constant (\ref{eq:OrthogonalRelationsContinuum251}) diffusion parameters.
%

\subsubsection{Constant Diffusion Parameter}

We consider the diffusion system 
corresponding to the symmetric star topology with constant diffusion parameter (\ref{eq:OrthogonalRelationsContinuum251}), its solution (\ref{eq:FinalAnswerUniformWeights505}) and spectrum (\ref{eq:Eigenvalues68}).
%
%
According to \cite{Robustness2010Young} the robustness $(H)$ is defined as
\begin{equation}
    \nonumber
    \begin{gathered}
        H=\sqrt{\frac{1}{2\sum_{i=2}^{N}\text{Real}(\lambda_i)}}
    \end{gathered}
\end{equation}
where the number of eigenvalues $N$ can tend to infinity.
Considering the spectrum 
(\ref{eq:Eigenvalues68}) we have
\begin{equation}
    \nonumber
    \begin{gathered}
       \sum_{i=2}^{\infty}\frac{1}{\text{Real}(\lambda_i)}=\frac{4}{\hat{\Theta} \pi^2}\sum_{k=1}^{\infty}\frac{1}{(2k)^2}+\frac{4(p-1)}{\hat{\Theta} \pi^2}\sum_{k=0}^{\infty}\frac{1}{(2k+1)^2}
    \end{gathered}
\end{equation}
Now considering the Riemann zeta-functions 
\cite{RiemannZetaRef1,RiemannZetaRef3}
given as below
\begin{equation}
    \nonumber
    \begin{gathered}
       \zeta(s)=\sum_{n=1}^{\infty}\frac{1}{n^s},\quad \zeta(2)=\frac{\pi^2}{6},
    \end{gathered}
\end{equation}
we obtain the following,
\begin{equation}
    \nonumber
    \begin{gathered}
       \sum_{k=1}^{\infty}\frac{1}{(2k)^2}=\frac{\pi^2}{24},\quad  \sum_{k=0}^{\infty}\frac{1}{(2k+1)^2}=\frac{\pi^2}{8}.
    \end{gathered}
\end{equation}
Thus the robustness of the system with constant diffusion parameter 
is as below,
\begin{equation}
    \label{eq:RobustnessFinalFormulaConstantDiffusionParameter2448}
    \begin{gathered}
       H = \frac{1}{2}\sqrt{\frac{3p-2}{3\hat{\Theta}}}
    \end{gathered}
\end{equation}

\subsubsection{Variable Diffusion Parameter}

\begin{figure}
  \centering
     \includegraphics[width=100mm]{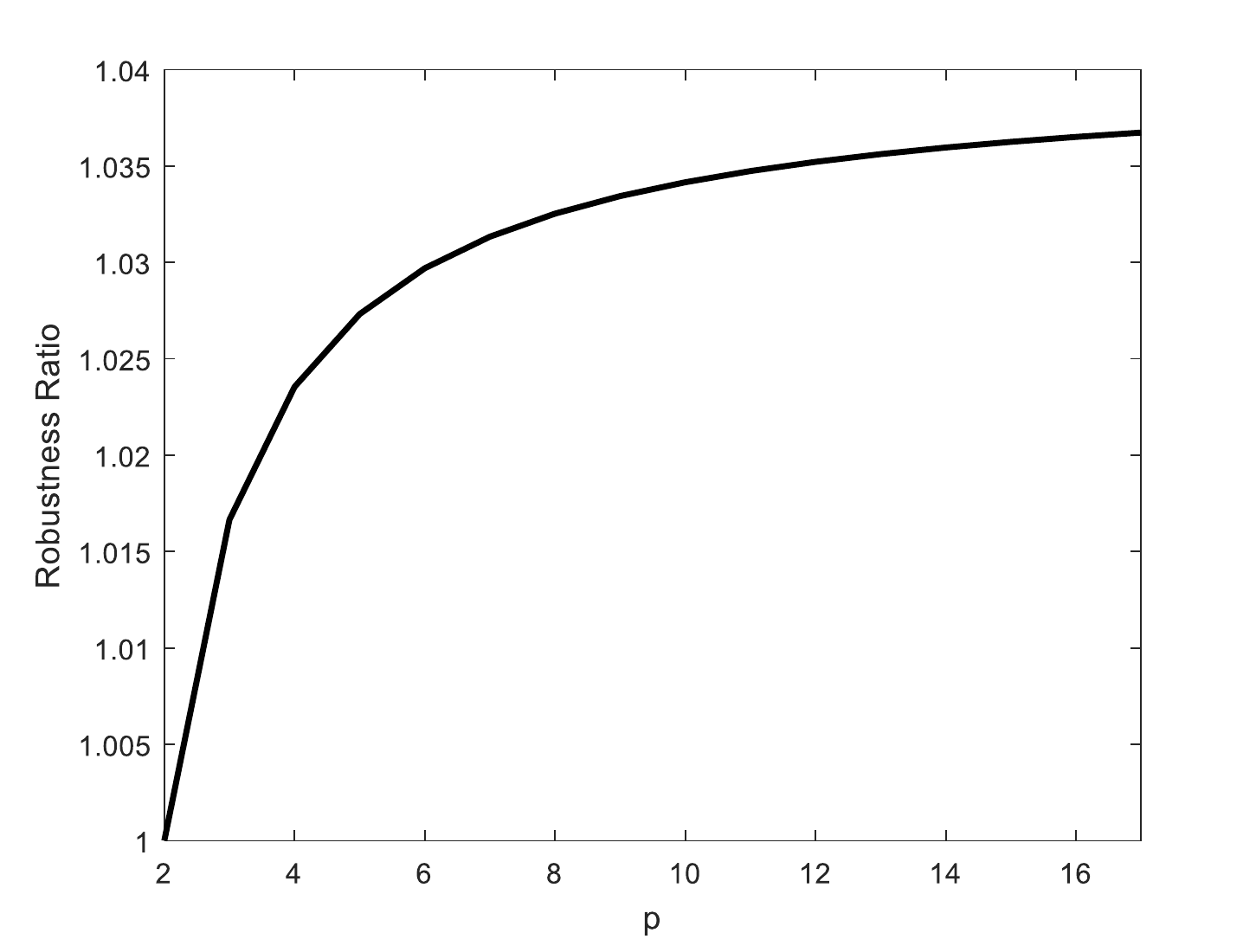}
  \caption{Ratio of the robustness obtained for constant diffusion parameter (\ref{eq:RobustnessFinalFormulaConstantDiffusionParameter2448}) to the robustness obtained for variable diffusion parameter (\ref{eq:RobustnessFinalFormulaVariableDiffusionParameter2494}).}
  \label{fig:FigureRobustnessRatio}
\end{figure}

Considering the diffusion system 
corresponding to the symmetric star topology with variable diffusion parameter (\ref{eq:NewStateUpdateEqContinuum130}) and its eigenvalues (\ref{eq:EigenvalueNonUniform123}),
the following can be concluded
\begin{equation}
    \label{eq:OrthogonalRelations373_2698}
    \begin{gathered}
        \sum_{i=2}^{\infty}\frac{1}{\text{Real}(\lambda_i)}=\frac{2}{3 \hat{\Theta} }\sum_{k=1}^{\infty}\frac{1}{2k(2k+1)}+\frac{2(p-1)}{3 \hat{\Theta} }\sum_{k=0}^{\infty}\frac{1}{2(k+1)(2k+1)}.
    \end{gathered}
\end{equation}
Now considering the alternating harmonic series \cite{RiemannZetaRef3} given as below,
\begin{equation}
    \label{eq:OrthogonalRelations373_2706}
    \begin{gathered}
      \sum_{n=1}^{\infty}\frac{(-1)^{n+1}}{n}=\ln2,
    \end{gathered}
\end{equation}
we can state the following,
\begin{subequations}
    \label{eq:OrthogonalRelations373_2714}
    \begin{align}
      \sum_{k=1}^{\infty}\frac{1}{2k(2k+1)}=\sum_{k=1}^{\infty}(\frac{1}{2k}-\frac{1}{2k+1})=1-      \ln2,  \\
      \sum_{k=0}^{\infty}\frac{1}{2(k+1)(2k+1)}=\sum_{k=0}^{\infty}(\frac{1}{2k+1}-\frac{1}{2(k+1)})= \ln2.
    \end{align}
\end{subequations}
Thus the Robustness of the system with variable diffusion parameter is as below,
\begin{equation}
    \label{eq:RobustnessFinalFormulaVariableDiffusionParameter2494}
    \begin{gathered}
       H=\sqrt{\frac{1+(p-2)\ln2}{3\hat{\Theta}}}.
    \end{gathered}
\end{equation}
In figure \ref{fig:FigureRobustnessRatio}, we have plotted the ratio between the robustness of the diffusion system 
obtained for constant diffusion parameter (\ref{eq:RobustnessFinalFormulaConstantDiffusionParameter2448}) to that obtained for variable diffusion parameter (\ref{eq:RobustnessFinalFormulaVariableDiffusionParameter2494}) in terms of $p$ (the number of branches in the symmetric star topology).
From figure \ref{fig:FigureRobustnessRatio}, it is obvious that the diffusion system 
with variable diffusion parameter is more robust.
%
%

\section{Conclusion}
\label{Conclusion}

In this paper, spatially-discrete continuous-time distributed consensus algorithm is approximated with a spatially-continuous system governed by linear partial differential equations which form a system of diffusion equations, referred to as the diffusion system.
This is done in an effort to exploit the algebraic structure of the system and enhance its convergence rate to its equilibrium state (i.e. consensus state).
Previously in the literature \cite{ItalyDiffusion2015}, the diffusion parameter for the obtained diffusion system 
has been assumed to be constant.
This assumption is equivalent to assigning constant weight to all edges of the underlying graph in the continuous-time distributed consensus algorithm.
Here, we have relaxed this assumption and assumed a spatially-variable diffusion parameter.
This has enabled us to address this optimization problem in a more general setting.
Optimizing the obtained diffusion system, 
we have shown that the convergence rate of the system (towards its equilibrium) has improved compared to the case with constant diffusion parameter.
In our solution, the diffusion system 
is achieved by approximating the state update equations of the spatially-discrete and continuous-time consensus algorithm in the continuum limit.
As a result of this approximation, the network is divided into two parts, the spatially-continuous path branches and the lattice core which connects these branches at one end.
Based on the optimal weights obtained in \cite{SaberQConsensusContinuous} for spatially-discrete continuous-time consensus algorithm, we have formulated the optimal weights for the lattice core and the spatially-variable diffusion parameter for the path branches.
Interestingly, the obtained results for the lattice core and the path bars are individually optimal, but in general, for a network with an arbitrary topology, the obtained results are suboptimal.
We have shown that the Symmetric star topology is an exception and for this topology the obtained results are globally optimal.
Moreover, we have validated the obtained results for the symmetric star topology via variational method, and we have shown that compared to the constant diffusion parameter, variable diffusion parameter improves the robustness of the diffusion system  
with symmetric star topology.

Future studies will be focused on solving other problems of coupled PDEs using the variational method,
This will lay the foundation for optimizing the discretized version of these problems.

\begin{appendices}

\section{Proof of Lemma \ref{lemma1}}
\label{sec:Proof1}
%
%
%
In this section, proof of Lemma \ref{lemma1} is provided.
\begin{proof}

For the first element of the state vector $\boldsymbol{Q}(\xi,t)$ corresponding to $\lambda_{1}(\boldsymbol{L}_{w})$, the boundary and the initial conditions are as below,
\begin{equation}
    \label{eq:FirstElementBoundary1239}
    \begin{aligned}
        \boldsymbol{\eta}_{1}^{T} \times \frac{\partial}{\partial \xi} \boldsymbol{Q}(\xi,t)|_{\xi=0} = 0,
    \end{aligned}
\end{equation}
\begin{equation}
    \label{eq:FirstElementInitial1246}
    \begin{aligned}
        \boldsymbol{\eta}_{1}^{T} \times \boldsymbol{Q}(\xi,0)  =  \frac{1}{\sqrt{N}} \boldsymbol{1}^{T} \times \boldsymbol{Q}(\xi,0).
    \end{aligned}
\end{equation}
This is due to the fact that $\lambda_{1}(\boldsymbol{L}_{w}) = 0$ and $\boldsymbol{\eta}_{1} = \frac{1}{\sqrt{N}} \boldsymbol{1}$.
The answer to this PDE is of the following form,
\begin{equation}
    \label{eq:FirstGeneralAnswer1256}
    \begin{aligned}
        \boldsymbol{\eta}_{1}^{T} \times \boldsymbol{Q}(\xi,t)  =  A_{1}  \cdot  e^{-\mu_{1} \cdot t} \cdot \cos{\left( \sqrt{ \mu_{1}/\hat{\Theta} } \cdot \xi \right)}.
    \end{aligned}
\end{equation}
Substituting (\ref{eq:FirstGeneralAnswer1256}) in the boundary conditions (\ref{eq:BoundaryConditions240a}) and (\ref{eq:FirstElementBoundary1239}), we obtain $\sin{\left( \sqrt{\frac{\mu_1}{\hat{\Theta}}} \right)} = 0$ which results in the following
\begin{equation}
    \label{eq:FirstMUAnswer413}
    \begin{aligned}
        \mu_{1,n_{1}}  =   n_{1}^{2} \cdot \pi^{2} \cdot \hat{\Theta}  \quad \text{for} \quad n_{1} = 0, \ldots, \infty.
    \end{aligned}
\end{equation}
Thus $\boldsymbol{\eta}_{1}^{T} \times \boldsymbol{Q}(\xi,t)$ can be written as below,
\begin{equation}
    \label{eq:FirstsummationForm422Uniform}
    \begin{aligned}
        \boldsymbol{\eta}_{1}^{T} \times \boldsymbol{Q}(\xi,t)
        =
        \sum_{n_{1}=0}^{\infty} { A_{1,n_{1}} \cdot e^{- \mu_{1,n_1} \cdot t } \cdot \cos{ \left( n_{1} \cdot \pi \cdot \xi \right) } }.
    \end{aligned}
\end{equation}
Substituting (\ref{eq:FirstMUAnswer413}) in (\ref{eq:FirstGeneralAnswer1256}) for $t=0$ (i.e. the initial condition (\ref{eq:FirstElementInitial1246})) we have
\begin{equation}
    \label{eq:FirstFourierSeries423}
    \begin{aligned}
        \frac{1}{\sqrt{N}} \cdot \boldsymbol{1}^{T} \times \boldsymbol{Q}(\xi,0)  =  \sum_{ n_{1} = 0 }^{\infty} {  A_{1,n_{1}} \cdot \cos{\left( n_{1} \cdot \pi \cdot \xi \right)}  }.
    \end{aligned}
\end{equation}
From the Fourier series expansion of $\frac{1}{\sqrt{N}} \cdot \boldsymbol{1}^{T} \times \boldsymbol{Q}(\xi,0)$ the $A_{1,n_{1}}$ coefficients  are obtained as below,
\begin{subequations}
    \label{eq:FirstACoefficients431}
    \begin{gather}
       A_{1,0} = \frac{1}{\sqrt{N}} \cdot \int_{0}^{1} {  \boldsymbol{1}^{T} \times \boldsymbol{Q}(\xi,0) d\xi  }  ,  \label{eq:FirstACoefficients431a} \\ 
       A_{1,n_{1}} = \frac{2}{\sqrt{N}} \cdot \int_{0}^{1} {  \left(  \boldsymbol{1}^{T} \times \boldsymbol{Q}(\xi,0) \right)  \cdot \cos{\left( n \cdot \pi \cdot \xi \right)} d\xi  },  \quad \text{for} \quad n_{1} \geq 1  \label{eq:FirstACoefficients431b}
    \end{gather}
\end{subequations}
Note that the results in (\ref{eq:FirstACoefficients431}) are based on the relation $\int_{0}^{1}{\cos{\left( n \cdot \pi \cdot \xi \right)} \cdot \cos{\left( m \cdot \pi \cdot \xi \right)}} = \frac{1}{2}\delta_{n,m} \cdot \left( \delta_{n,0}  +  1  \right)$.
Regarding other components of the state vector $\boldsymbol{Q}(\xi,t)$ (i.e. $\boldsymbol{\eta}_{k}^{T} \times \boldsymbol{Q}(\xi,t)$ for $k \geq 2$) the answer to the PDE defined by (\ref{eq:DiffusionPDEStateUpdate229}) and (\ref{eq:BoundaryConditions404}) is of the following form
%
\begin{equation}
    \label{eq:KthGeneralAnswerUniform1304}
    \begin{aligned}
        \boldsymbol{\eta}_{k}^{T} \times \boldsymbol{Q}(\xi,t)  =  A_{k} \cdot e^{ -\mu_{k} \cdot t } \cdot \cos{\left(  \sqrt{ \frac { \mu_{k} } { \hat{\Theta} } } \cdot (1-\xi)  \right)}.
    \end{aligned}
\end{equation}
By substituting (\ref{eq:KthGeneralAnswerUniform1304}) in the boundary condition (\ref{eq:BoundaryConditions240b}), equation (\ref{eq:KthMuAnswerUniform558}) is obtained. 
Thus $\boldsymbol{\eta}_{k}^{T} \times \boldsymbol{Q}(\xi,t)$ can be written as below,
\begin{equation}
    \label{eq:XSum459Uniform}
    \begin{aligned}
        \boldsymbol{\eta}_{k}^{T} \times \boldsymbol{Q}(\xi,t)  =  \sum_{n_{k} = 1}^{\infty} {  A_{k,n_{k}} \cdot e^{ -\mu_{k,n_{k}}\cdot t } \cdot \cos{ \left( \sqrt{\frac{\mu_{k,n_{k}}}{\hat{\Theta}}} \cdot (1-\xi) \right) }  }
    \end{aligned}
\end{equation}
The coefficients $A_{k,n_{k}}$ for $k=2,\ldots,N$ are obtained from the Fourier series expansion of $\boldsymbol{\eta}_{k}^{T} \times \boldsymbol{Q}(\xi,t)$ as below,
\begin{equation}
    \label{eq:KthACoefficientUniform1332}
    \begin{aligned}
        A_{k,n_{k}}  =  \left(    \frac { 2 \lambda_{k} }  { \lambda_{k} + \sin^{2}{\left( \frac{\mu_{k,n_{k}}}{\hat{\Theta}} \right)} }    \right)  \cdot  \int_{0}^{1} {  \left( \boldsymbol{\eta}_{k}^{T} \times \boldsymbol{Q}_{0}(\xi) \right) \cdot \cos{ \left(  \sqrt{ \frac { \mu_{k,n_{k}} } { \hat{\Theta} } } \cdot (1-\xi)  \right) } \cdot d\xi  }.
    \end{aligned}
\end{equation}
The results in (\ref{eq:KthACoefficientUniform1332}) are based on the following relation
\begin{equation}
    \label{eq:KthOrthogonalRelationUniform1340}
    \begin{aligned}
        \int_{0}^{1} { \cos{ \left( \sqrt{ \frac{\mu_{k,n_{k}}}{\hat{\Theta}} } \cdot \xi  \right) } \cdot \cos{ \left( \sqrt{ \frac{\mu_{k,m_{k}}}{\hat{\Theta}} } \cdot \xi \right) } \cdot d\xi   }
        =
        \frac{1}{2}\left( 1 + \frac{ \sin^{2}{ \left( \sqrt{ \frac{\mu_{k,n_{k}}}{\hat{\Theta}} } \right) } }  {\lambda_{k}} \right) \cdot \delta_{n_{k},m_{k}}.
    \end{aligned}
\end{equation}
To achieve (\ref{eq:KthOrthogonalRelationUniform1340}) and thus (\ref{eq:KthACoefficientUniform1332}), we define $y_{k,n_{k}} = \sqrt{ \frac{\mu_{k,n_{k}}}{\hat{\Theta}} }$.
It is obvious that $ \frac { \partial^{2} } { \partial\xi^{2} } \cos{ \left( y_{k,n_{k}} \cdot \xi \right) }  +  y_{k,n_{k}}^{2} \cdot \cos{ \left( y_{k,n_{k}} \cdot \xi \right) }  =  0  $.
Thus we can write the following
\begin{equation}
    \label{eq:Uniform1352}
    \begin{aligned}
        &\frac{\partial}{\partial \xi}
        \left(
        \cos{ \left( y_{k,m_{k}} \cdot \xi \right) } \cdot \frac{\partial}{\partial \xi} \cos{ \left( y_{k,n_{k}} \cdot \xi \right) }
        -
        \cos{ \left( y_{k,n_{k}} \cdot \xi \right) } \cdot \frac{\partial}{\partial \xi} \cos{ \left( y_{k,m_{k}} \cdot \xi \right) }
        \right)
        +  \\
        & \qquad \qquad \qquad \qquad \qquad \qquad \qquad \qquad
        \left(  y_{k,n_{k}}^{2}  -  y_{k,m_{k}}^{2}  \right)  \cdot  \cos{ \left( y_{k,n_{k}}\cdot \xi \right) } \cdot \cos{ \left( y_{k,m_{k}}\cdot \xi \right) }
        =  0.
    \end{aligned}
\end{equation}
Integrating (\ref{eq:Uniform1352}) from $0$ to $1$ we obtain the following
\begin{equation}
    \label{eq:Uniform1372}
    \begin{aligned}
        &y_{k,m_{k}}  \cdot  \cos{\left( y_{k,n_{k}} \right)}  \cdot  \sin{\left( y_{k,m_{k}} \right)}
        -
        y_{k,n_{k}}  \cdot  \cos{\left( y_{k,m_{k}} \right)}  \cdot  \sin{\left( y_{k,n_{k}} \right)}
        + \\
        & \qquad\qquad \qquad\qquad \qquad\qquad
        \left(   y_{k,n_{k}}^{2}  -  y_{k,m_{k}}^{2}   \right)  \cdot  \int_{0}^{1} {  \cos{\left( y_{k,n_{k}} \cdot \xi \right)}  \cdot    \cos{\left( y_{k,m_{k}} \cdot \xi \right)}  d\xi }.  
    \end{aligned}
\end{equation}
For $n_{k} \neq m_{k}$, the expression in (\ref{eq:Uniform1372}) is equal to zero and for $n_{k} = m_{k}$ it is equal to $  \frac{1}{2}  +  \frac{\sin^{2}{y_{k,n_{k}}}}{2\lambda_{k}}  $.
This implies the result in (\ref{eq:KthOrthogonalRelationUniform1340}).
Using the fact that
$\sum_{k=1}^{N} { \boldsymbol{\eta}_{k} \times \boldsymbol{\eta}_{k}^{T} }  =  \boldsymbol{I}$,
the states vector $\boldsymbol{Q}(\xi,t)$ can be written as
$\boldsymbol{Q}(\xi,t)  =  \boldsymbol{I} \times \boldsymbol{Q}(\xi,t)  =   \sum_{k=1}^{N} { \left( \boldsymbol{\eta}_{k}^{T} \times \boldsymbol{Q}(\xi,t) \right) \cdot \boldsymbol{\eta}_{k} } $. 
By substituting $\boldsymbol{\eta}_{k}^{T} \times \boldsymbol{Q}(\xi,t)$ from (\ref{eq:FirstsummationForm422Uniform})  and (\ref{eq:XSum459Uniform}) in the above expression, we obtain (\ref{eq:539Uniform}) for the state vector $\boldsymbol{Q}(\xi,t)$.

\end{proof}

\section{Proof of Lemma \ref{lemma2}}
\label{sec:Proof2}

This section provides the proof of Lemma \ref{lemma2}.

\begin{proof}
Similar to the case of constant diffusion parameter, for the first element of the state vector $\boldsymbol{Q}(\xi,t)$ corresponding to $\lambda_{1}(\boldsymbol{L}_{w})$, the boundary and the initial conditions are as in (\ref{eq:FirstElementBoundary1239}) and (\ref{eq:FirstElementInitial1246}).
For the first element of the state vector $\boldsymbol{Q}(\xi,t)$ (i.e. $\boldsymbol{\eta}_{1}^{T} \times \boldsymbol{Q}(\xi,t)$), the answer to PDE defined by the diffusion equation (\ref{eq:StateUpdateEquationsContinuumVariableThetaFinalForm1117}) is of the following form,
\begin{equation}
    \label{eq:FirstGeneralAnswer404}
    \begin{aligned}
        \boldsymbol{\eta}_{1}^{T} \times \boldsymbol{Q}(\xi,t)  = \sum_{n_{1}=0}^{\infty}A_{1,2n_1}  \cdot  e^{- \mu_{1,n_1}^{'} \cdot t} \cdot P_{2n_1}{\left(  \xi \right)},
    \end{aligned}
\end{equation}
where $\mu_{1,n_1}^{'} = \frac{3\hat{\Theta}}{2}2n_1(2n_1+1)$ and
\begin{equation}
    \label{eq:FirstGeneralAnswer1741}
    \begin{aligned}
              A_{1,2n_1}(0)=(4n_1+1)\int_0^1 { P_{2n_1}(\xi) \cdot \left( \boldsymbol{\eta}_{1}^{T} \times \boldsymbol{Q}(\xi,0) \right)  \cdot  d\xi }.
   \end{aligned}
\end{equation}
$P_{2n_1}{\left(  \xi \right)}$ is the Legendre polynomial of order $2n_1$ (as explained in Appendix \ref{sec:Legendre}).
Regarding other components of the state vector $\boldsymbol{Q}(\xi,t)$ (i.e. $\boldsymbol{\eta}_{k}^{T} \times \boldsymbol{Q}(\xi,t)$ for $k \geq 2$) the answer to the PDE defined by (\ref{eq:StateUpdateEquationsContinuumVariableThetaFinalForm1117}) and (\ref{eq:StateUpdateEquationsContinuumVariableThetaFinalForm1170}) is of the following form
\begin{equation}
    \label{eq:KthGeneralAnswer441}
    \begin{aligned}
        \boldsymbol{\eta}_{k}^{T} \times \boldsymbol{Q}(\xi,t)  = \sum_{n_{k} = 1}^{\infty} {  A_{k,n_{k}} \cdot e^{ - \mu_{k,n_k}^{'} \cdot t }} \cdot P_{\nu_{k,n_k}}{\left(  \xi  \right)},
    \end{aligned}
\end{equation}
where $\mu_{k,n_k}^{'}  =  \frac{3\hat{\Theta}}{2}\nu_{k,n_k}(\nu_{k,n_k}+1)$ and the parameters $\nu_{k,n_k}$ for $k=2,\ldots,N$ and $n_{k} = 1,\ldots,\infty$ are obtained from the roots of (\ref{eq:nuPolynomial1259}).
%
%
$P_{\nu_{k,n_k}}$ is the Legendre polynomial of order $\nu_{k,n_k}$ and the coefficients $A_{k,n_{k}}$ for $k=2,\ldots,N$ in (\ref{eq:KthGeneralAnswer441}) are obtained from the expansion of $\boldsymbol{\eta}_{k}^{T} \times \boldsymbol{Q}(\xi,t)$ as below,
\begin{equation}
    \label{eq:KthACoefficient470}
    \begin{aligned}
        A_{k,n_{k}}  =  \left(    \frac { 1 }  { \int_{0}^{1}\left( P_{\nu_{k,n_k}}{\left(  \xi  \right)}\right)^2 \cdot d\xi} \right)  \cdot  \int_{0}^{1} {  \left( \boldsymbol{\eta}_{k}^{T} \times \boldsymbol{Q}_{0}(\xi) \right)} \cdot P_{\nu_{k,n_k}}{\left(  \xi  \right)} \cdot d\xi .
    \end{aligned}
\end{equation}
The results in (\ref{eq:KthACoefficient470}) are obtained using the following relation
\begin{equation}
    \label{eq:KthOrthogonalRelation478}
    \begin{aligned}
        \int_{0}^{1} P_{\nu_{k,n_k}}{\left(  \xi  \right)} \cdot P_{\nu_{k,m_k}}{\left(  \xi  \right)} \cdot d\xi
        =
        \left(\int_{0}^{1}\left( P_{\nu_{k,n_k}}{\left(  \xi  \right)}\right)^2d\xi\right) \cdot \delta_{n_{k},m_{k}}.
    \end{aligned}
\end{equation}
In the following we explain how (\ref{eq:KthOrthogonalRelation478}) 
is obtained based on the Legendre equation.
Legendre equation can be written as below,
%
\begin{equation}
    \label{eq:LegendreEquation1821}
    \begin{gathered}
        \frac { d }{d \xi}( (1-\xi^2) \frac {d P_{\nu_{k,m_k}}{\left(  \xi  \right)}}{d\xi} )  + \nu_{k,m_k} \cdot (1+\nu_{k,m_k}) \cdot P_{\nu_{k,m_k}}{\left(  \xi  \right)}  =  0
    \end{gathered}
\end{equation}
%
Based on (\ref{eq:LegendreEquation1821}) we have
\begin{equation}
    \label{eq:1861}
    \begin{aligned}
        &\frac{d}{d \xi}
        \left(
      (1-\xi^2) \cdot P_{\nu_{k,m_k}}{\left(  \xi  \right)} \cdot \frac{d P_{\nu_{k,n_k}}{\left(  \xi  \right)}}{d \xi}
        -
        (1-\xi^2) \cdot P_{\nu_{k,n_k}}{\left(  \xi  \right)} \cdot \frac{d P_{\nu_{k,m_k}}{\left(  \xi  \right)}}{d \xi}
        \right)
        +  \\
        & \qquad
        \left( \nu_{k,m_k} \cdot (1+\nu_{k,m_k})  - \nu_{k,n_k} \cdot (1+\nu_{k,n_k}) \right)  \cdot P_{\nu_{k,m_k}}{\left(  \xi  \right)} \cdot P_{\nu_{k,n_k}}{\left(  \xi  \right)}
        =  0.
    \end{aligned}
\end{equation}
Integrating (\ref{eq:1861}) from $0$ to $1$ and using the boundary condition $\frac{d P_{\nu_{k,m_k}}{\left(  \xi  \right)}}{d \xi} \big{|}_{\xi=0} = \lambda_2 \cdot P_{\nu_{k,m_k}}{\left(  0  \right)}$ we obtain the following,
%
\begin{equation}
    \label{eq:509}
    \begin{aligned}
        &\lambda_2\left( P_{\nu_{k,m_k}}{\left(  0  \right)}   \cdot  P_{\nu_{k,n_k}}{\left(  0  \right)}
        -
        P_{\nu_{k,n_k}}{\left(  0  \right)}   \cdot  P_{\nu_{k,m_k}}{\left(  0  \right)} \right)
        + \\
        & \qquad\qquad \qquad\qquad \qquad
        \left( \nu_{k,m_k}(1+\nu_{k,m_k})  - \nu_{k,n_k}(1+\nu_{k,n_k}) \right)  \cdot \int_{0}^1 P_{\nu_{k,m_k}}{\left(  \xi  \right)} \cdot P_{\nu_{k,n_k}}{\left(  \xi  \right)}=0.  
    \end{aligned}
\end{equation}
For $n_{k} \neq m_{k}$, the expression in (\ref{eq:509}) is equal to zero.
This implies the result in (\ref{eq:KthOrthogonalRelation478}).
Using the fact that
$\sum_{k=1}^{N} { \boldsymbol{\eta}_{k} \times \boldsymbol{\eta}_{k}^{T} }  =  \boldsymbol{I}$,
the states vector $\boldsymbol{Q}(\xi,t)$ can be written as
$\boldsymbol{Q}(\xi,t)  =  \boldsymbol{I} \times \boldsymbol{Q}(\xi,t)  =   \sum_{k=1}^{N} { \left( \boldsymbol{\eta}_{k}^{T} \times \boldsymbol{Q}(\xi,t) \right) \cdot \boldsymbol{\eta}_{k} } $. 
By substituting $\boldsymbol{\eta}_{k}^{T} \times \boldsymbol{Q}(\xi,t)$  from (\ref{eq:FirstGeneralAnswer404}) and (\ref{eq:KthGeneralAnswer441}) in the above expression, we obtain (\ref{eq:FinalResultVariableTheta1236}) for the state vector $\boldsymbol{Q}(\xi,t)$.

\end{proof}

\section{Legendre Differential Equation and its solutions}
\label{sec:Legendre}
%
%
%
Legendre's differential equation \cite{LegendreBookHilbert1965,LegendreBook1968} is defined as below,
\begin{equation}
    \label{eq:1932}
    \begin{gathered}
         \frac{d}{d \xi} \left( (1-\xi^2)  \frac{d}{d \xi}  \boldsymbol{y}  \right) + n (n+1) \boldsymbol{y}=0. 
    \end{gathered}
\end{equation}
%
For nonnegative integer $n$, the solution of the Legendre's differential equation is as below,
\begin{equation}
    \label{eq:1943}
    \begin{aligned}
         P_n(\xi) = \Sigma_{k=0}^{\lfloor{\frac{n}{2}}\rfloor}\frac{(-1)^k (2n-2k)!}{2^n k!(n-k)!(n-2k)!}\xi^{n-2k}.
    \end{aligned}
\end{equation}
$P_n(\xi)$ is referred to as the Legendre polynomial of order $n$ and it is finite for $-1 \leq \xi \leq 1$.
Another formulation for Legendre polynomials is the Rodrigues' formulation given as below,
\begin{equation}
    \label{eq:1968}
    \begin{aligned}
         P_{n}(\xi) = \frac{(-1)^{n}}{2^{n}n!}\frac{d^n}{d\xi^{n}}(1-n^2)^{n}
    \end{aligned}
\end{equation}
%
Few Legendre polynomials of lower orders are as below,
\begin{equation}
    \label{eq:1952}
    \begin{aligned}
         P_0(\xi)=1,\quad P_1(\xi)=\xi, \quad P_2(\xi)=\frac{1}{2}(3\xi^2-1),\quad P_3(\xi)=\frac{1}{2}(5\xi^3-3\xi).
    \end{aligned}
\end{equation}
For the marginal values, we also have,
\begin{equation}
    \begin{aligned}
         P_{n}(1) = 1,\quad P_{n}(-1)=(-1)^{n}, \quad P_{2n+1}(0)=0,\quad P_{2n}(0) = \frac{(-1)^{n}(2n)!}{2^{2n}n!}.
    \end{aligned}
\end{equation}
The following orthonormality condition holds between Legendre polynomials,
\begin{equation}
    \label{eq:LegendreOrthogonality1993}
    \begin{aligned}
         \int _{-1}^{1}P_{n}(\xi)P_m(\xi)d\xi=\frac{2}{2n+1}\delta_{n m}
    \end{aligned}
\end{equation}
The second solution of Legendre differential equation which is independent from $P_{n}(\xi)$ is called the Legendre function of second kind and it is defined as below,
\begin{equation}
    \label{eq:484}
    \begin{aligned}
        Q_{n}(\xi)
        =
        \frac{1}{2}P_{n}(\xi)\ln\frac{1+\xi}{1-\xi}
        -
        \sum_{k=0}^{[\frac{n-1}{2}]}\frac{2n-4k-1}{(2k+1)(n-k)}P_{n-2k-1}(\xi),
    \end{aligned}
\end{equation}
where
\begin{equation}
    \label{eq:2009}
    \begin{aligned}
        Q_{0}(\xi)=\frac{1}{2}\ln\frac{1+\xi}{1-\xi},
    \end{aligned}
\end{equation}
and the function $\left[\frac{n-1}{2}\right]$ is defined as below,
\begin{equation}
    \nonumber
    \begin{gathered}
    \left[\frac{n-1}{2}\right]=
               \begin{cases}
            \frac{n-1}{2} \quad \text {if $n$ is odd}  \\
            \frac{n-2}{2} \quad \text {if $n$ is even}
        \end{cases}
    \end{gathered}
\end{equation}

\section{Hypergeometric Function}
\label{sec:Hypergeometric}

Hypergeometric function \cite{LegendreBook1968,LegendreBookHilbert1965}  
is defined as below,
\begin{equation}
    \nonumber
    \begin{gathered}
 {}_2F_1(\alpha, \beta, \gamma, \xi)=\Sigma_{r=0}^{\infty}\frac{(\alpha)_r(\beta)_r\xi^r}{(\gamma)_rr!},
    \end{gathered}
\end{equation}
where  $(\alpha)_r$ is defined  as $(\alpha)_r = \alpha(\alpha + 1)\cdots (\alpha + r - 1)$.
The derivative of hypergeometric function is also a hypergeometric function and it can be written as below, 
\begin{equation}
    \nonumber
    \begin{gathered}
\frac{d}{d\xi}({}_2F_1(\alpha, \beta, \gamma, \xi))=\frac{\alpha\beta}{\gamma} {}_2F_1(\alpha+1, \beta+1, \gamma+1, \xi)
    \end{gathered}
\end{equation}
The following relation holds between the hypergeometric functions and the Legendre functions and their derivative,
\begin{equation}
    \nonumber
    \begin{gathered}
 P_{\nu}(\xi)={}_2F_1(-\nu, \nu+1, 1, \frac{1-\xi}{2})\\\frac{d}{d\xi}P_{\nu}(\xi) = \frac{\nu(\nu+1)}{2} \cdot {}_2F_1(-\nu+1, \nu+2, 2, \frac{1-\xi}{2})
    \end{gathered}
\end{equation}

\section{Sturm Comparison Theorem}
\label{SturmLiouvilleAppendix}
Here, we prove that $\mu_{1,2}>\mu_{2,1}$. 
Based on (\ref{eq:DiffusionPDESturmLiouvilleBetaNonZeroVariational148}) and (\ref{eq:DiffusionPDESturmLiouvilleBetaZeroVariational158}),
for the solutions of $ \Phi_{1,2}(\xi)$ and $\Phi_{2,1}(\xi)$ with corresponding eigenvalues $\mu_{1,2}$ and $\mu_{2,1}$, respectively, we have
\begin{equation}
    \label{eq:Eq1AppendixSturmLiouville325}
    \begin{aligned}
        &\frac{d}{d \xi} \left(\Theta(\xi) \left(\Phi_{1,2}(\xi) \frac{d\Phi_{2,1}(\xi)}{d \xi}-\Phi_{2,1}(\xi) \frac{d\Phi_{1,2}(\xi)}{d \xi} \right)\right)=
        \\
        &\qquad\qquad
        \Phi_{1,2}(\xi) \frac{d}{d \xi} \left(\Theta(\xi)  \frac{d}{d \xi}\Phi_{2,1}(\xi) \right)-\Phi_{2,1}(\xi) \frac{d}{d \xi} \left(\Theta(\xi)  \frac{d}{d \xi}\Phi_{1,2}(\xi)\right)=
        \\
        &\qquad\qquad\qquad\qquad\qquad\qquad\qquad\qquad\qquad\qquad\qquad\qquad
        (\mu_{1,2}- \mu_{2,1})\Phi_{1,2}(\xi)\Phi_{2,1}(\xi)
    \end{aligned}
\end{equation}
Now $\Phi_{1,2}(\xi)$ has a single root in the interval $(0,1)$ and we denote it by $x_1$ and $ \Phi_{2,1}(\xi)$ is non-zero in the interval  $(0,1)$ and we can assume that it is positive  but $\Phi_{2,1}(0)=0 $ 
\cite{LegendreBook1968,LegendreBookHilbert1965}.
Without loss of generality we can assume that $ \Phi_{1,2}(\xi)$ is positive in the interval $[0,x_1]$.
Assuming $\mu_{1,2}< \mu_{2,1}$,  
%
%
%
it can be concluded that the function
$\left(\Theta(\xi) \left(\Phi_{1,2}(\xi) \frac{d\Phi_{2,1}(\xi)}{d \xi}-\Phi_{2,1}(\xi) \frac{d\Phi_{1,2}(\xi)}{d \xi} \right)\right)$ is a nondecreasing function in the interval $[0,x_1]$.
But on the other hand, we have
$\left(\Theta(\xi) \left(\Phi_{1,2}(\xi) \frac{d\Phi_{2,1}(\xi)}{d \xi} \right. \right.$ $\left.\left.\left.-\Phi_{2,1}(\xi) \frac{d\Phi_{1,2}(\xi)}{d \xi} \right)\right)\right|_{\xi=0} = \Theta(\xi)\Phi_{1,2}(\xi) \frac{d\Phi_{2,1}(\xi)}{d \xi}|_{\xi=0}>0 $
and
$\left(\Theta(\xi) \left(\Phi_{1,2}(\xi) \frac{d\Phi_{2,1}(\xi)}{d \xi}-\Phi_{2,1}(\xi)\right.\right.$ $\left.\left.\left.\frac{d\Phi_{1,2}(\xi)}{d \xi} \right)\right)\right|_{\xi=x_{1}}$  $=$  $\Phi_{2,1}(\xi)\cdot$ $\frac{d\Phi_{1,2}(\xi)}{d \xi}$ $|_{\xi=x_1}<0$
(since $\frac{\Phi_{1,2}(\xi)}{d \xi}|_{\xi=x_1}<0 $), 
which is in contradiction with the conclusion that the function
$\left(\Theta(\xi) \left(\Phi_{1,2}(\xi) \frac{d\Phi_{2,1}(\xi)}{d \xi}-\Phi_{2,1}(\xi) \frac{d\Phi_{1,2}(\xi)}{d \xi} \right)\right)$ is a nondecreasing function in the interval $[0,x_1]$.
Therefore, the assumption $\mu_{1,2}< \mu_{2,1}$ is not correct and we have $\mu_{1,2}>\mu_{2,1}$.

\end{appendices}

\bibliographystyle{ieeetran}
\bibliography{acmsmall-sample-bibfile}


\end{document}